%% file: Article.tex
\documentclass{aa}
\usepackage[flushleft]{threeparttable}
\usepackage{supertabular}
\usepackage{txfonts}
\usepackage{xcolor}
\usepackage{lscape}
\usepackage{natbib}
\usepackage{enumitem}   
\usepackage{graphicx}
\usepackage{amssymb}
\usepackage[colorlinks=true,allcolors=blue]{hyperref}

\begin{document}

   \title{A new indirect approach for oxygen abundance determinations across a wide range of physical conditions}

    \titlerunning{A new indirect method using the $SO_{23}$ parameter}
    \authorrunning{Zamora et al.}
    
   \author{S.~Zamora\inst{\ref{inst:SNS}}
          \and
          Asier Castrillo \inst{\ref{inst:OAN}}
          \and
          A.~I. D\'{i}az\inst{\ref{inst:UAM}\ref{inst:CIAFF}}
          \and
          Marta Gavil\'{a}n \inst{\ref{inst:UAM}}
          }

   \institute{
              Scuola Normale Superiore, Piazza dei Cavalieri 7, I-56126 Pisa, Italy \label{inst:SNS}
              \and
              Observatorio Astron\'{o}mico Nacional (IGN), C/ Alfonso XII 3, E-28014 Madrid, Spain \label{inst:OAN}
              \and
               Departamento de F\'{i}sica Te\'{o}rica, Universidad Aut\'{o}noma de Madrid, Madrid 28049, Spain \label{inst:UAM}           
               \and
               CIAFF, Universidad Autónoma de Madrid, 28049 Madrid, Spain \label{inst:CIAFF}                
}


\abstract{In this work we present a new approach to derive HII region metal abundances based on sulfur emission lines. We select M101 as benchmark galaxy to test our new methodology along the wide metallicity gradient of this huge face-on star-forming spiral. Based on previous studies, we compile a sample of 151 HII regions and compare their properties base on the galaxy radial profile. The new methodology is based on the derivation of the electron temperature of the double ionised sulfur, either by the direct measurement of the [SIII] auroral line or, at supersolar metalicities, from the $SO_{23}$ parameter. This new indirect method is able to extend the abundance gradient up to the very centre of the M101 galaxy (0.04 $R_e$). This new methodology mitigates systematic biases in direct abundance determinations based on auroral lines, while opening up new opportunities to improve commonly used strong line abundance calibrations in the intermediate to high metallicity regime.}


\maketitle

\section{Introduction} \label{sec_intro}

The gas-phase metallicity of galaxies is a fundamental property that provides key insights into their chemical enrichment and evolutionary history. The metal content distribution is essential for understanding the interplay between gas accretion, star formation, and feedback processes that regulate galaxy evolution \citep{Maiolino2019}. However, accurate determinations of chemical abundances remains challenging because of the wide range of physical conditions exhibit throughout galaxies as well as the limitations inherent to the different abundance diagnostics.

The derivation of chemical abundances in galaxies relies primarily on the analysis of emission lines from HII regions, clouds of gas ionised by hot, massive stars that can be readily identified across galactic discs, even at large distances. Oxygen has traditionally been used as the main tracer of gas-phase metallicity, as it is the most abundant element after hydrogen and helium. Since it is predominantly produced by massive stars and rapidly released into the interstellar medium (ISM), its abundance is commonly associated with the present-day chemical composition of galaxies. Young stars typically have sufficiently high effective temperatures to ionise significant amounts of $O^{+}$ and $O^{++}$ in the surrounding nebulae. Collisional excitation of these ions by free electrons, followed by radiative de-excitation, gives rise to the forbidden emission lines $[OII]\lambda\lambda$3727,3729 \AA\ and $[OIII]\lambda\lambda$4959,5007 \AA, which are readily accessible in the optical spectral range \citep{Aller1984}. The free electron distribution can be characterised by the electron temperature ($T_e$), which governs the collisional excitation rates of the ionic species present in the nebula. In this context, an accurate determination of the electron temperature derived from the auroral lines allows the oxygen ionic abundances to be derived directly from the measured emission line intensities \citep[i.e.][]{Castellanos2002,Hagele2008}.

This approach, commonly known as the “direct method”, is widely considered as one of the most reliable methods for determining oxygen abundances. However, despite its reliability, the direct method is subject to significant selection effects and biases, particularly in the central regions of galaxies, which provide key anchor points for establishing abundance gradients. First, the exponential dependence of collisionally excited line emissivities on $T_e$ means that the observed line intensities are strongly weighted towards warmer parts of the nebular ionisation structure \citep{Shields1978}. This effect is particularly important for oxygen, for which the temperature derived from the $[OIII]$ lines can exceed the actual ionic temperature of $O^{++}$, $T_e[OIII]>T(O^{++})$ \citep[][]{Stasinska2005}. At high metallicities, most of the oxygen is in the form of $O^+$ (87--95\%; \citealt{Diaz_Zamora2022}) then this effect cannot be considered important. Nonetheless, $T_e[OII]$ is estimated from $T_e[OIII]$, and then the temperature of the $O^+$ emitting gas may also be overestimated, particularly in the coolest regions. This leads to an underestimation of $O^+/H^+$, the dominant contributor to the total $O/H$ abundance. Consequently, oxygen abundances derived using $T_{e}$-based methods may be systematically biased towards low metallicities, potentially by substantial factors \citep[see also][]{Stasinska1978,Stasinska2002,Zamora2026}. Second, in high metallicity regions, auroral lines are often difficult or impossible to detect  since their emissivities decrease rapidly at low electron temperatures. For example, the auroral $[OIII]\lambda$4363 \AA\ line is detectable only for $T_{e}\gtrsim 7000 K$ \citet{Bresolin2008}.

These observational limitations motivate the use of the cooling properties of oxygen emission lines as an alternative approach to calibrating gas-phase metallicities. Several combinations of emission lines have been proposed as abundance calibrators. Among them, one of the most widely used is the $R_{23}$ (or $O_{23}$) parameter, originally proposed by \citet{Pagel1979}, which is defined as the sum of the intensities of the $[OII]\lambda\lambda3727,3729$ and $[OIII]\lambda\lambda4959,5007$ emission lines, normalised to the intensity of $H\beta$ ($\lambda4861$). A comprehensive list of the abundance indices proposed to date can be found in \citet{Stasinska2010}. These indices can be empirically calibrated against abundances derived using the direct method, although this approach is generally restricted to low-to-moderate metallicity regions, where auroral lines can be detected. At higher metallicities, their calibration relies primarily on photoionisation model predictions. However, strong-line methods are subject to systematic uncertainties and are not free from biases. HII regions are, at least, a three-parameter family, their emission line spectrum being controlled by the shape of the ionising radiation, the ionisation parameter and the chemical composition of the gas. Therefore, in principle, they should be used only for HII regions that show characteristics similar to those of the calibrating sample objects and the calibrations should not be used out of their range of validity.

Using oxygen as abundance tracer and IFS techniques \citet{Sanchez2024} found an  appreciably drop in abundances in the inner regions of 26 of the galaxies with HII regions analysed from the CALIFA sample. These galaxies are of early type and do not show evidence of the presence of bars or interactions with neighbours. In a later work, \citet{Sanchez-Menguiano2018} studied a sample of 102 spiral galaxies observed with VLT/MUSE, comprising a total of 14345 HII regions, to derive their oxygen abundance distribution using the $O3N2$ calibrator. This study yielded evidence for oxygen abundance drops in about a third of the galaxy sample. Interestingly enough, according to the authors, the most massive systems present the deepest drops which may cast some doubts about the methodology employed to derive oxygen abundances in the high metallicity regime.

An alternative to the limitations associated with using oxygen as a reference element for determining gas-phase metallicity gradients is to consider alternative elements whose auroral lines remain detectable at lower electron temperatures, thereby extending the applicability of the direct method to higher metallicities. Sulphur is a particularly suitable candidate for this purpose. It is predominantly produced in massive stars, and its nucleosynthetic yields are expected to track those of oxygen. Hence, sulphur can also be used as an abundance tracer, providing similar information on the chemical enrichment and evolutionary history of galaxies. The auroral emission line of $[SIII]$ at $\lambda $6312 are analogues to that of $[OIII]$ at $\lambda$4363. Furthermore, in the case of sulphur, due to the lower excitation energies involved, the line emissivities are less dependent on $T_e$ which allows the line detection even at over-solar abundances \citep{Diaz2007, Diaz_Zamora2022}. For example, \citet{Bresolin2008} estimated that the auroral $[OIII]\lambda$4363 \AA\ line is only detectable for $T_{e}\gtrsim 7000 K$, while $[SIII]\lambda$6312 \AA\ can be detected at temperatures down to approximately 5000 K.

In this work, we present an alternative methodology for the derivation of oxygen abundances across spiral discs using sulphur as electron temperature tracer and paying special attention to the high metallicity regime. We have applied this methodology to the case of M101 (NGC5457). It is located at a distance of 6.52 Mpc \citep[][corresponding to a linear scale of 31.6 pc/arcsec]{Beaton2019} and has an effective radius ($R_{e}$) of 197.6 arcsec \citep[][]{Berg2020}, corresponding to a physical radius of 6.26 kpc at the adopted distance. This galaxy contains a large number of very luminous Giant Extragalactic H II Regions (GEHRs) and it has been extensively studied using spectroscopic \citep[e.g.,][]{Kennicutt_Garnett1996,Kennicutt2003,Li2013,Croxall2016,Hu2018} and photometric\citep{Cedres2004,Garner2024,Thaina-Batista2025} techniques. Therefore, it provides an excellent benchmark for testing the new proposed methodology. 

The sample selection and new observations are presented in Section \ref{data}. Section \ref{sec:method} describes the new methodology, while the results and discussion are presented in Sections \ref{sec:res} and \ref{discusion}, respectively. Finally, the main conclusions are summarised in Section \ref{sec:conclusions}. The physical conditions, including electron temperatures and chemical abundances, were calculated using PyNeb \citep{Luridiana2015} , based on a five-level statistical equilibrium model and adopting the atomic coefficients listed in Table 3 of \citet{Fernandez2018}. Our final sample as well as the the complete set of derived properties along this work can be found in electronic format.

\section{Sample and observations} \label{data}

\subsection{Data compilation} \label{compilation}

M101 (NGC5457) is an almost face-on local spiral galaxy at recorded distances between 6.4 and 7.4 Mpc. The galaxy is undergoing tidal interactions with two late-type dwarf galaxies: NGC 5474 and NGC 5477, within a nearby galaxy group \citep{Linden2022}. It has a large population of HII regions more than a thousand of which were catalogued by \citet{Hodge1990}. An appreciable number of these HII regions are bright enough to allow a complete spectroscopic analysis. For the present work we collected the largest number of observations of HII regions available that include measurements of the following emission lines: $[OII]\lambda\lambda$3737,29 \AA, $[OIII]\lambda\lambda$4959,5007 \AA, $[NII]\lambda\lambda 6548,84$ \AA, $[SII]\lambda\lambda$6717,31 \AA, and $[SIII]\lambda$9069 \AA, although, in many cases, the lines of $[OIII]\lambda$4363 \AA\ ,$[SIII]\lambda$6312 \AA\ and $[SIII]\lambda$9532 \AA\ are also included. Auroral line measurements with S/N < 2 were excluded from the analysis. We applied homogeneous and consistent procedures to reliably determine the abundances in each region. The data included in our compilation come from the following sources: \citet{Kennicutt_Garnett1996}, \citet{Kennicutt2003}, \citet{Bresolin2007}, and \citet{Croxall2016}, resulting in a total of 131 HII regions. 

\subsection{SDSS observations} \label{SDSS}

Additionally, spectra of 20 HII regions from the list in \citet{pilyugin2007} were extracted from Data Release 5 (DR5) of the Sloan Digital SkySurvey (SDSS). According to the authors, each of the HII regions selected shows the $[OIII]\lambda$4363 \AA\ auroral line at high metallicities (i.e. an oxygen abundance of $12 + log(O/H) \geq 8.2$). 

In this case, we did not directly use the measurements reported by the authors, as they noted that the extinction derived from the H$\alpha$/H$\beta$ ratio was negative. After visual inspection of the spectra, we found that 8 of them show saturation in the strongest emission lines, including H$\alpha$. We therefore measured all the emission lines in these spectra following the procedure described in \citet{marta2025} \citep[see also][]{ngc7742}. Then, we used the available HeI emission lines in each spectrum, normalised to $HeI$ $\lambda6678$ \AA, to derive the dust extinction following the methodology proposed by \citet{Zamora2022}. We adopted the Galactic extinction law of \citet{reddening}, with $R_V=2.97$, and the theoretical Balmer and helium ratios assuming $n_e=100 cm^{-3}$ and $T_e=10^4$ K \citep[][]{cloudy}. In both cases, we assumed a simple screen distribution of the dust and the same extinction for the emission lines and the stellar continuum. The reddening coefficient $c(H\beta)$ was derived through a linear regression using all the available HI and HeI emission lines in each spectrum. In the cases in which $[OIII]\lambda5007$ was saturated, we estimated the flux from $[OIII]\lambda4959$ using the theoretical intensity ratio of 3 between the two emission lines. These measurements can be found in Appendix \ref{ap:red}.

\section{New abundance methodology}\label{sec:method}

\subsection{The direct method}

When the sulphur auroral line is detectable, the electron temperature of $S^{++}$, $T_e[SIII]$, is derived from the ratio of the nebular to auroral [SIII] line intensities, $R_{S3} = ([SIII]\lambda9069 + [SIII]\lambda9532)/[SIII]\lambda6312$. Given the broad range of metallicities covered by the radial gradients, we adopt the expression from \citet{Diaz_Zamora2022}, which is valid for $5000 \leq T_e[SIII] \leq 25000$ K and exhibits only a weak dependence on electron density \citep{Perez-montero2017,Perez-Montero2005}. This choice is particularly important for the central, metal-rich regions of the galaxies, where the electron temperatures can fall below $8000$ K, outside the validity range of many other expressions available in the literature. In total, the sulphur auroral line is detected in 92 regions, corresponding to 61\% of the sample.

Similarly, when the auroral line is detectable, the electron temperature of $O^{++}$, $T_e[OIII]$, is derived using the ratio of the nebular to auroral [OIII] line intensities, $R_{O3} = ([OIII]\lambda4959 + [OIII]\lambda5007)/[OIII]\lambda4363$. The adopted expression is valid for $7000 \leq T_e[OIII] \leq 25000$ K \citep{Diaz_Zamora2022}. In total, the oxygen auroral line is detected in 81 regions (54\% of the sample).

\subsection{The indirect method}

In this work, we propose a new methodology, referred to as the ``indirect'' method, to address these two biases affecting the central, high-metallicity regions of abundance gradients: (i) the exponential dependence of collisionally excited line emissivities on $T_e$ means that the observed line intensities are strongly weighted towards warmer parts of the nebular ionisation structure \citep{Shields1978}; and (ii) the oxygen auroral line, $[OIII]\lambda$4363 \AA, is often difficult or impossible to detect in high metallicity regions ($T_{e}$ lower than $7000 K$) because their emissivities decrease rapidly at low electron temperatures \citep{Bresolin2008}. Our new approach estimates $T_e[SIII]$ from strong emission lines, avoiding the selection effects associated with the detection of auroral lines, and uses it to derive direct chemical abundances. By estimating the oxygen and sulphur electron temperatures from $T_e[SIII]$, we also mitigate the bias arising from the overestimation of $T_e[OII]$ when inferred from $T_e[OIII]$. 

Several strong line calibrations have been proposed in the literature to estimate chemical abundances in ionised nebulae, based on individual elements or combinations of emission lines from different species \citep[e.g.][]{Pagel1979,Denicolo2002,Stasinska2006}. Among these, the $O_{23}$ parameter is one of the most widely used indicators of oxygen abundance. Its main advantage is that it combines strong [OII] and [OIII] emission lines, making it applicable to a broad range of HII regions, even when auroral lines are too faint to be detected. However, its calibration is double-valued: the same $O_{23}$ value can correspond to either the lower- or upper-metallicity branch, requiring an additional diagnostic to break this degeneracy. Moreover, at high metallicities, the parameter becomes less sensitive to abundance variations, limiting its ability to trace the inner regions of spiral galaxies. Another widely used strong line calibrator is the $S_{23}$ parameter \citep{Diaz2000}, which combines [SII] and [SIII] emission lines. Like $O_{23}$, it is double-valued, but its turnover occurs at over-solar metallicities. 

The ratio of $S_{23}$ and $O_{23}$ has therefore been proposed as a reliable abundance indicator, also for high-metallicity HII regions \citep[see][]{Perez-Montero2005}. It was defined as \citep{Diaz2000}:
\begin{equation}
SO_{23} = \frac{S_{23}}{O_{23}} =
\frac{[SII]\lambda\lambda6717,6731 + [SIII]\lambda\lambda9069,9532}
{[OII]\lambda\lambda3727,3729 + [OIII]\lambda\lambda4959,5007}
\end{equation}

This parameter provides a single-valued metallicity estimate because, when both parameters lie on their upper-metallicity branches, their complementary behaviour compensates for the trend of each individual calibrator. 

\citet{Diaz2007} used this parameter to derive an empirical calibration that, for the first time, enabled the electron temperature of $[SIII]$ to be estimated from the $SO_{23}$ parameter at over-solar metallicities:
\begin{equation}
T_e[SIII] = 0.596 - 0.283\times log (SO_{23}) + 0.199 \times log( SO_{23})^2
\end{equation}
where $T_e[SIII]$ is in units of K. This calibration is valid for $-1.0 < log(SO_{23}) < 0.5$.

\begin{figure}[t!]
\centering
\includegraphics[width=0.42\textwidth]{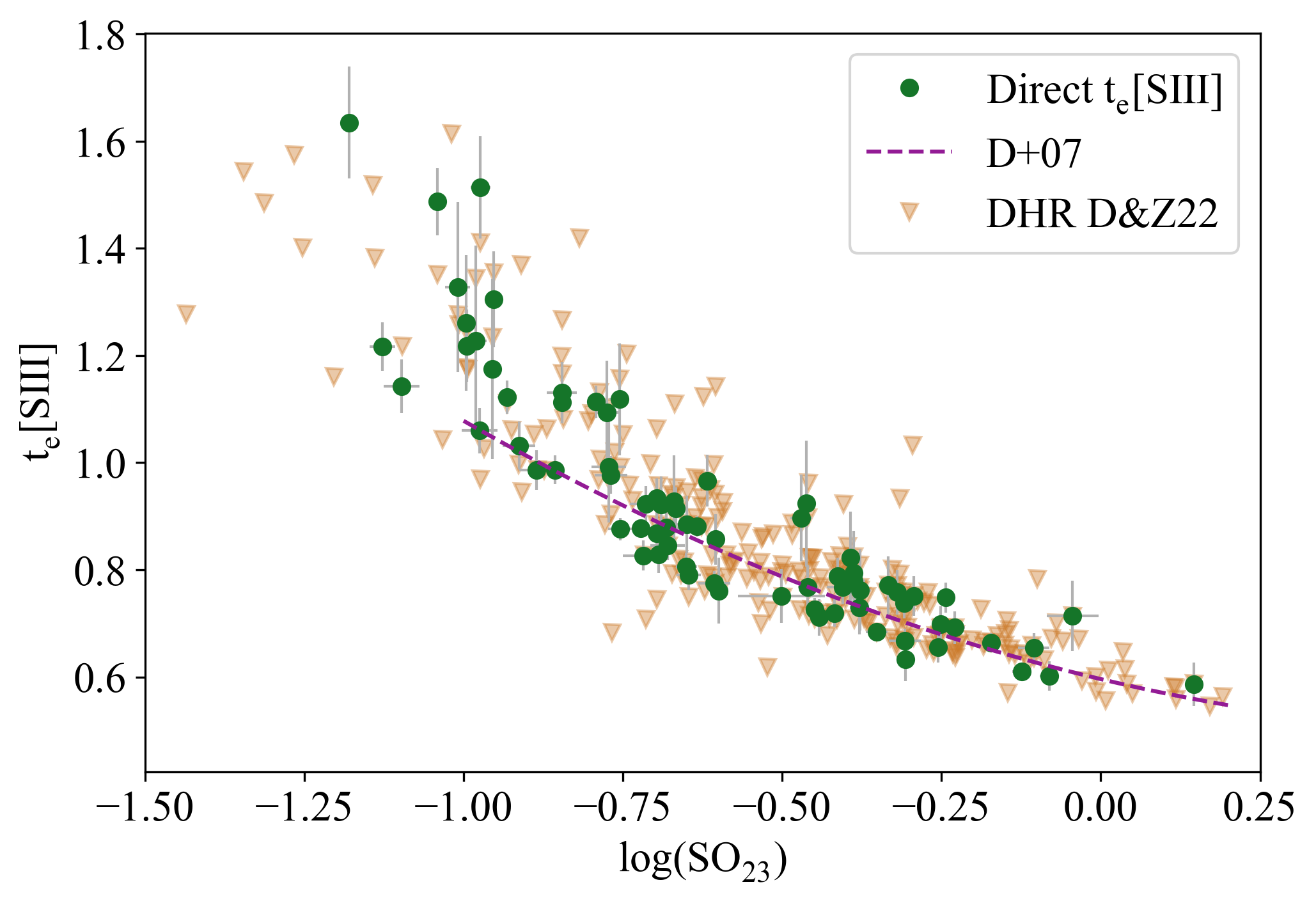}
\caption{Empirical calibration of the [SIII] electron temperature as a function of the abundance parameter $SO_{23}$ defined by \citet[][D+07]{Diaz2007} together with the direct sulphut electron temperatures compiled in this work. The compilation of HII regions from \citet[][D\&Z22]{Diaz_Zamora2022} is shown as yellow triangles. Electron temperature is in units 10$^4$ K.}
\label{fig_so23}
\end{figure}
Figure \ref{fig_so23} shows the empirical calibration of $T_e[SIII]$ as a function of $SO_{23}$ from \citet{Diaz2007}, together with the direct measurements of the sulphur electron temperatures compiled in this work. The good agreement between the calibration and the observed temperatures supports the use of this relation to estimate $T_e[SIII]$ in our sample, allowing us to include in the abundance gradients those central regions where direct electron temperature measurements are unavailable.

\subsection{Ionisation structure}\label{sec_ion_structure}

Once $T_e[SIII]$ has been determined from either the auroral line, or the indirect method described above, deriving chemical abundances requires assumptions about the ionisation structure of the nebula, since different ions are associated with regions of different electron temperatures and the ionisation structure varies with metallicity.

For oxygen, the most common approach assumes a two-zone ionisation structure: a high-ionisation zone, where $O^{++}$ originates and the electron temperature is represented by $T_e[OIII]$, and a low-ionisation zone, where $O^+$ originates and the electron temperature is represented by $T_e[OII]$.
Recent studies have found tight relations between $T_e[SIII]$ and $T_e[NII]$, as well as between $T_e[NII]$ and $T_e[OII]$ \citep{Croxall2016,Berg2020,Bresolin2004}. Based on these relations, we assume $T_e[SIII] \simeq T_e[NII] \simeq T_e[OII]$ \citep{Garnett1992,Scholte2026}. This assumption is also supported by the fact that, independently of the metallicity of the ionised nebulae, most of the sulphur is in the form of $S^{++}$ \citep[e.g.][]{Garnett1992,Diaz_Zamora2022}.

Finally, the electron temperature of [OIII] is estimated either directly from the auroral line, $[OIII]\lambda$4363 \AA\ when available, or from $T_e[SIII]$. Several relations have been proposed in the literature, but they show significant discrepancies, particularly at low electron temperatures. For instance, the relation derived from photoionisation models by \citet{Garnett1992} differs from the empirical calibration proposed by \citet{Croxall2016}, with temperature differences exceeding 2000 K in the coolest regions where the auroral $[SIII]\lambda6312$ line is detected. \citet{Croxall2016} noted that differences in the atomic data alone cannot explain these discrepancies and that their empirical relation is significantly steeper than the one derived by \citet{Garnett1992}. 

\begin{figure}[t!]
\centering
\includegraphics[width=0.42\textwidth]{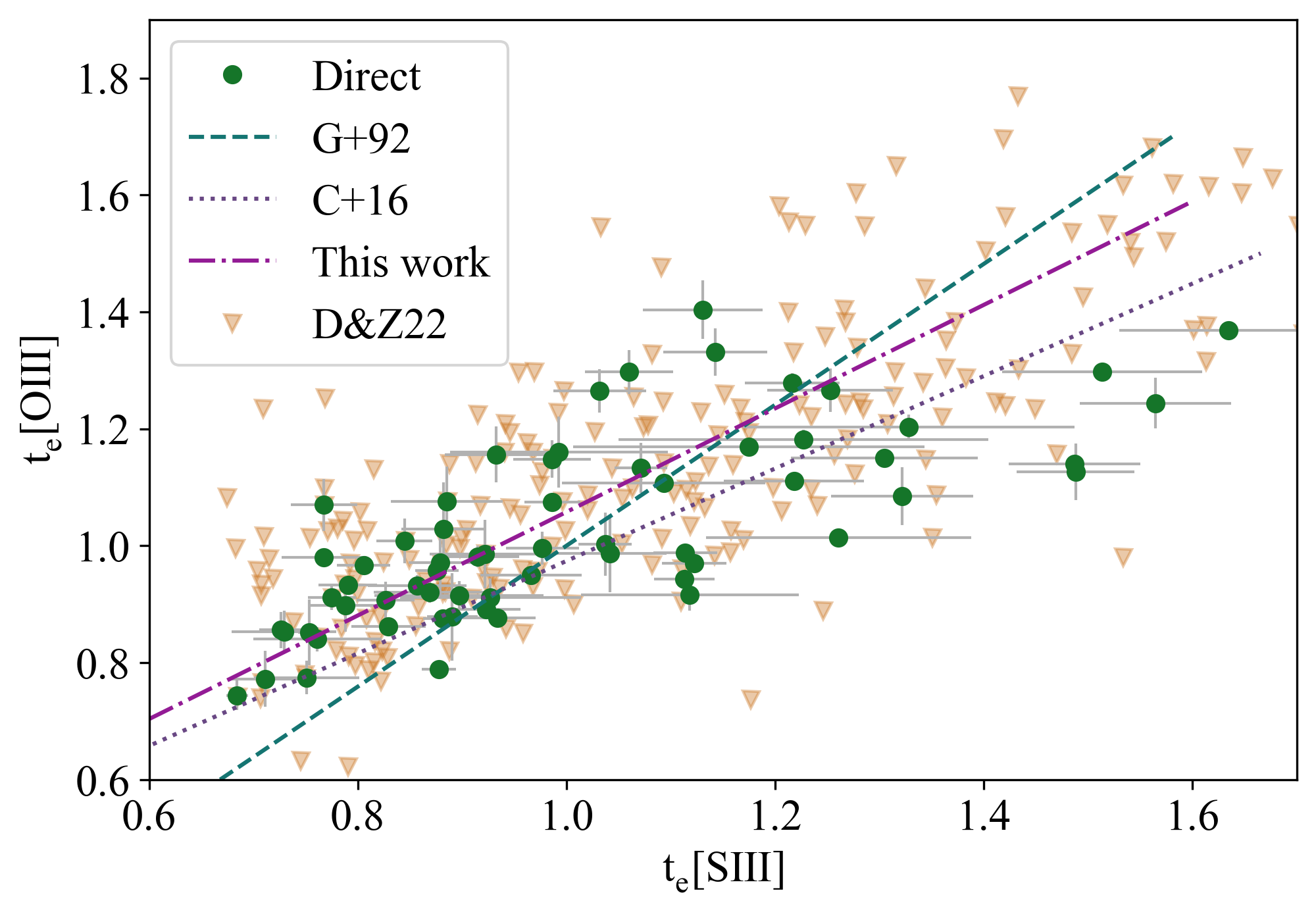}
\caption{Calibration between $T_e[OIII]$ and $T_e[SIII]$, compared with the relations from \citet[][G+92]{Garnett1992} and \citet[][C+16]{Croxall2016}. The yellow triangles represent the compilation by \citet[][D\&Z22]{Diaz_Zamora2022}, used to derive our new calibration, while the green circles show the direct $T_e$ measurements for both ions obtained in this work. Electron temperature is in units 10$^4$ K.}
\label{fig_tesiii_teoiii}
\end{figure}

Given this discrepancy and the importance of the coolest regions in our work for anchoring the abundance gradients, we derived a new relation between $T_e[OIII]$ and $T_e[SIII]$. To account for the broad range of physical conditions and metallicities relevant to our study, we used the data compiled by \citet{Diaz_Zamora2022}, which include a large sample of ionised nebulae, HII regions in disc galaxies, and HII galaxies. We fitted the data using orthogonal distance regression (ODR) with bootstrap resampling. The fit was performed with $T_e[OIII]$ as a function of $T_e[SIII]$, using inverse-variance weighting ($1/\mathrm \sigma^2$). The resulting relation is:
\begin{equation}
T_e[OIII] = (0.8853 \pm 0.0419)\times T_e[SIII] + (0.1725 \pm 0.0421)
\end{equation}
where $T_e[OIII]$ and $T_e[SIII]$ are in units of K. 

Figure \ref{fig_tesiii_teoiii} shows the relation between $T_e[SIII]$ and $T_e[OIII]$ derived in this work, compared with previous calibrations. We find a clear systematic offset between our temperature relation and that of \citep{Garnett1992}. This discrepancy is likely related to the different atomic data adopted in our abundance analysis compared to those used in the photoionisation models of Garnett (1992), which can introduce systematic offsets in the derived temperatures. Our temperature relation is also consistent with that of \citet{Croxall2016}, with the slopes agreeing within 1.0$\sigma$ and the intercepts being fully consistent (0.1$\sigma$). Despite this agreement, we adopt our new calibration because it is based on a large and diverse sample of HII regions and HII galaxies spanning a wide range of metallicities, making it well suited to the goals of the present work. Also, as shown in the figure, the direct ($T_e$) measurements for M101 from this work, particularly those at lower temperatures, show better agreement with our new calibration.

The uncertainties associated with the adopted ionisation structure and temperature relations have already been quantified by \citet{Diaz_Zamora2022}. We derived $T_e[OIII]$ from $T_e[SIII]$ rather than adopting a relation between $T_e[OII]$ and $T_e[OIII]$, since the latter is not straightforward. In particular, $T_e[OII]$ depends on the electron density, and the relation between the two temperatures is not universal, but varies with the physical conditions of the nebula \citep[see, e.g.,][]{Hagele2006,Izotov2006}. Moreover, the formal uncertainty in $T_e[OII]$ derived from such a relations accounts only for the propagation of the uncertainty in $T_e[OIII]$, without considering uncertainties associated with the adopted photoionisation models or their calibration. These limitations introduce additional uncertainties in the determination of $O^+/H^+$, which can be comparable to or greater than $O^{2+}/H^+$ in low-excitation nebulae ($T_e[SIII]<11000$ K), thereby significantly affecting the total $O/H$ abundance.

\begin{figure}[t!]
\centering
\includegraphics[width=0.42\textwidth]{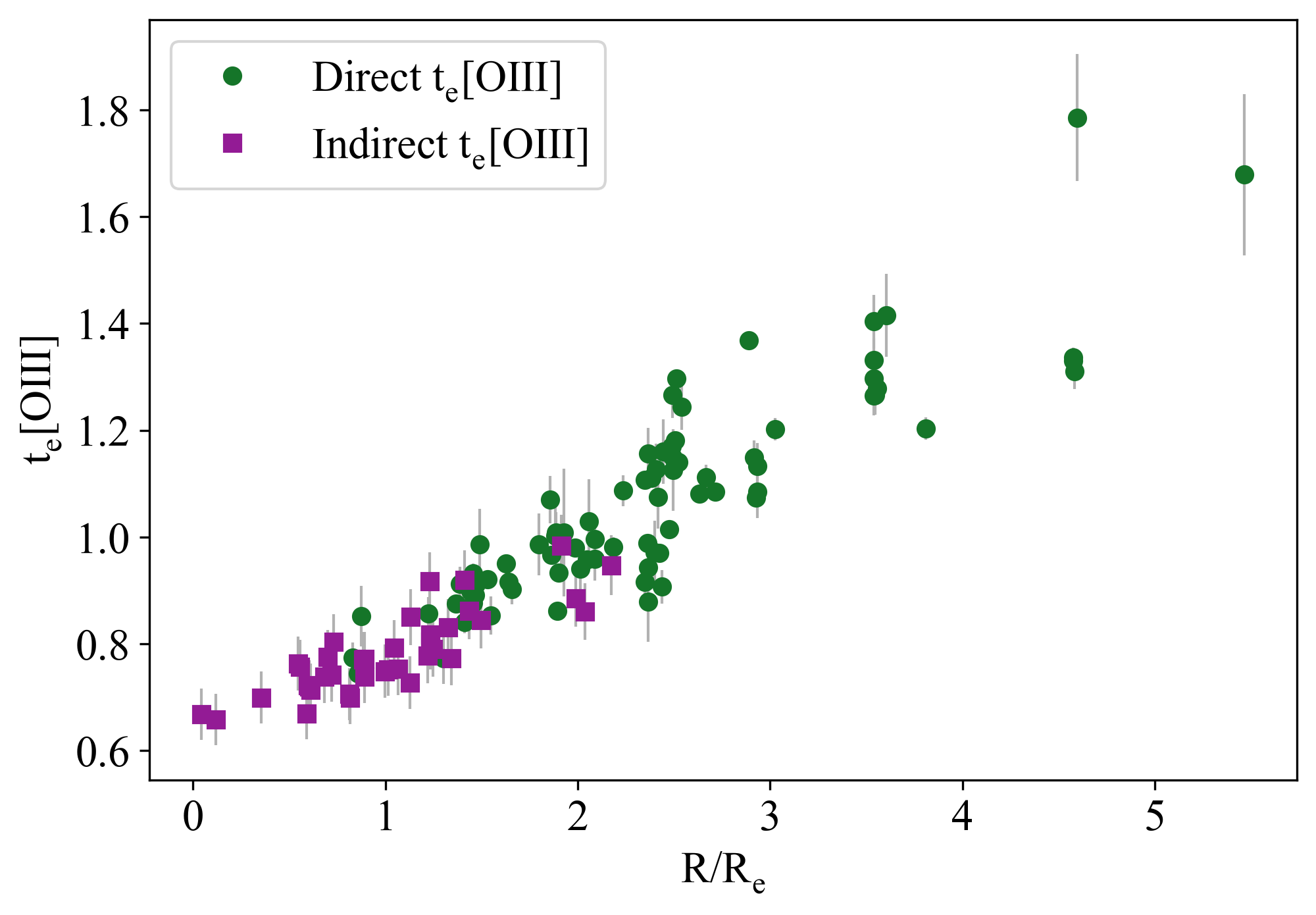}
\includegraphics[width=0.42\textwidth]{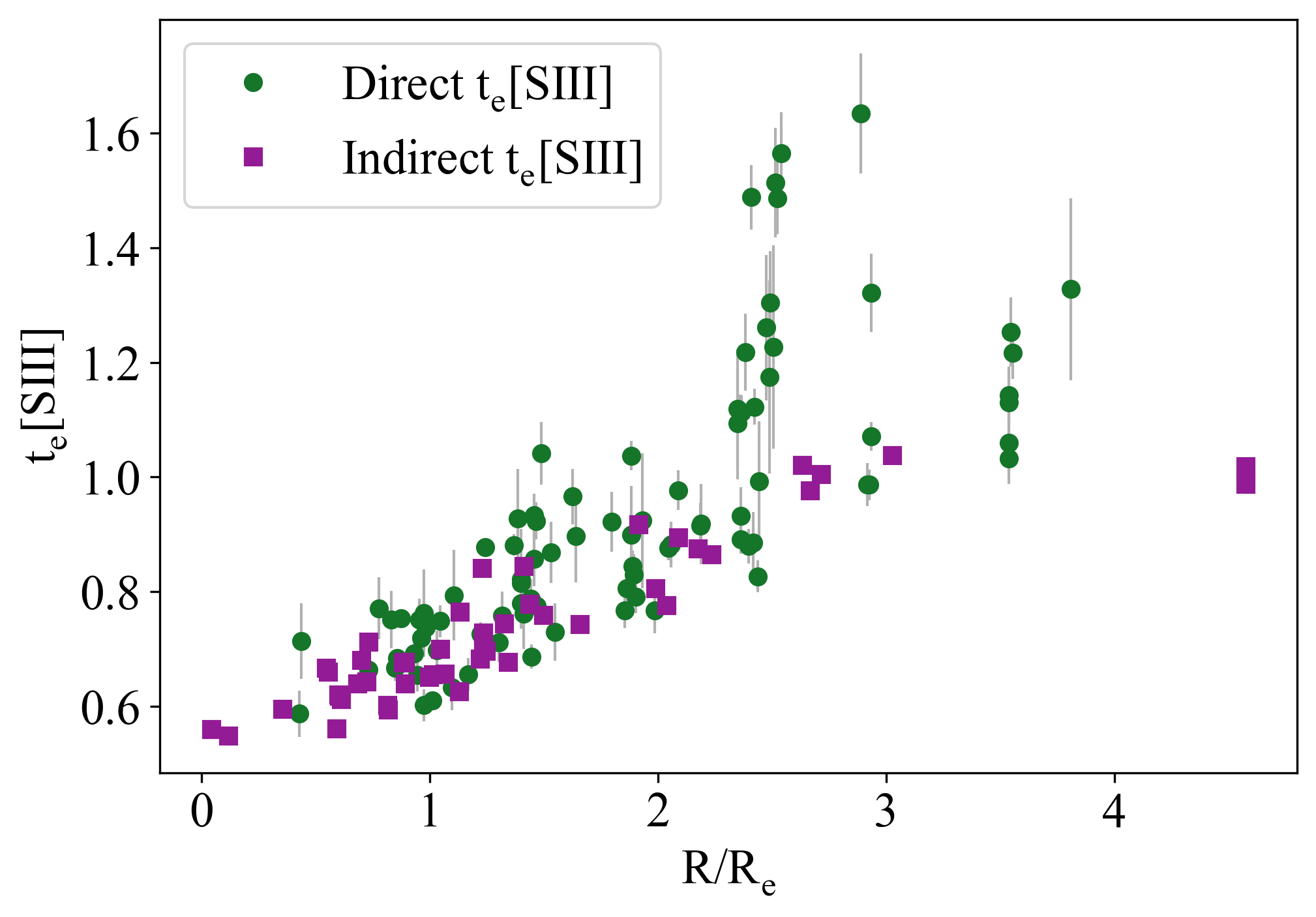}
\caption{Radial gradients of $T_e[OIII]$ (top) and $T_e[SIII]$ (bottom), derived using the direct (auroral lines) and the indirect method. Circles and triangles represent the direct and indirect determinations, respectively. The radial scale has been adjusted independently for each panel. Electron temperature is in units 10$^4$ K.}
\label{fig_te}
\end{figure}

Figure \ref{fig_te} shows the radial gradients of $T_e[OIII]$ and $T_e[SIII]$ across the galaxy disc. The electron temperatures are better constrained in the outer regions of the galaxy, where the metallicity is lower, as the auroral [OIII] line is not detected in the central regions ($R \lesssim 1.5 R_{eff}$), due to the exponential decrease in its emissivity at low electron temperatures. In contrast, the auroral [SIII] line remains detectable at smaller galactocentric distances, down to $R \sim 0.5 R_{eff}$. It is worth highlighting that the temperatures derived using the indirect method proposed in this work smoothly extend the trends traced by the direct measurements.

It should be noted that a subset of 26 regions present direct $T_e[SIII]$ determinations but no $[OIII]\lambda$4363 \AA\ auroral line detection. For simplicity and in order to improve the visual clarity, the conversion from $T_e[SIII]$ to $T_e[OIII]$ has not been applied in these cases. Consequently, the indirect $T_e[OIII]$ estimates shown here are derived exclusively from indirect $T_e[SIII]$ determinations. Therefore, the two panels of Figure \ref{fig_te} does not contain exactly the same regions and for this reason the two temperature estimates are compared separately.

\section{Results}\label{sec:res}

\subsection{Electron density}

The electron densities were determined from the $[SII]\lambda6717$ \AA/$[SII]\lambda6731$ \AA\ line ratio. In most cases, the measured ratios are below the critical value of this ratio, preventing a reliable determination of $n_e$. We therefore adopted a value of $n_e=$ 50 $cm^{-3}$ for these regions. Higher electron densities in the inner regions have been reported in several studies \citep{Kennicutt1989,Bresolin2004,Diaz2007,ngc7469}. However, most of the HII regions in our sample (see Figure \ref{fig:ne}) have electron densities characteristic of the low-density ISM, with the highest value being 410 $cm^{-3}$. We do not find any clear trend of electron density with galactocentric radius. Therefore, collisional de-excitation is not expected to significantly affect the strength of emission lines such as [OIII] in our sample, as the electron densities remain well below the densities at which collisional de-excitation becomes important for the relevant transitions \citep{Osterbrock2006}.

\subsection{Effective temperature}

The hardness of the ionising radiation field is difficult to constrain from commonly used excitation indices, such as $[OIII]\lambda$ 4363 \AA/H$\beta$, since they are sensitive to both nebular ionisation structure and metallicity. To overcome this limitation, \citet{Mathis1985} proposed the use of composite indices based on the ionisation states of different elements. In particular, the $O^+/O^{++}$ and $S^+/S^{++}$ ionic ratios can be combined to constrain the effective temperature, $T_{eff}$, with reduced dependence on metallicity and ionisation parameter. The relative populations of the different ionisation stages depend on the number of photons above their respective ionisation potentials. Therefore, a harder radiation field, which has a large fraction of high energy photons, set a different balance between the ionisation stages of O and S. This dependence allows the definition of the inverse excitation parameter $\eta$ \citep{Diaz87} and the use of its emission line version  $\eta'$, defined in \citet{Vilchez_pagel1988} for estimating the effective temperature of ionising stars, $T_{eff}$. Lower $\eta'$ values correspond to harder ionising spectra and higher effective temperatures. 

\begin{figure}[t!]
\centering
\includegraphics[width=0.42\textwidth]{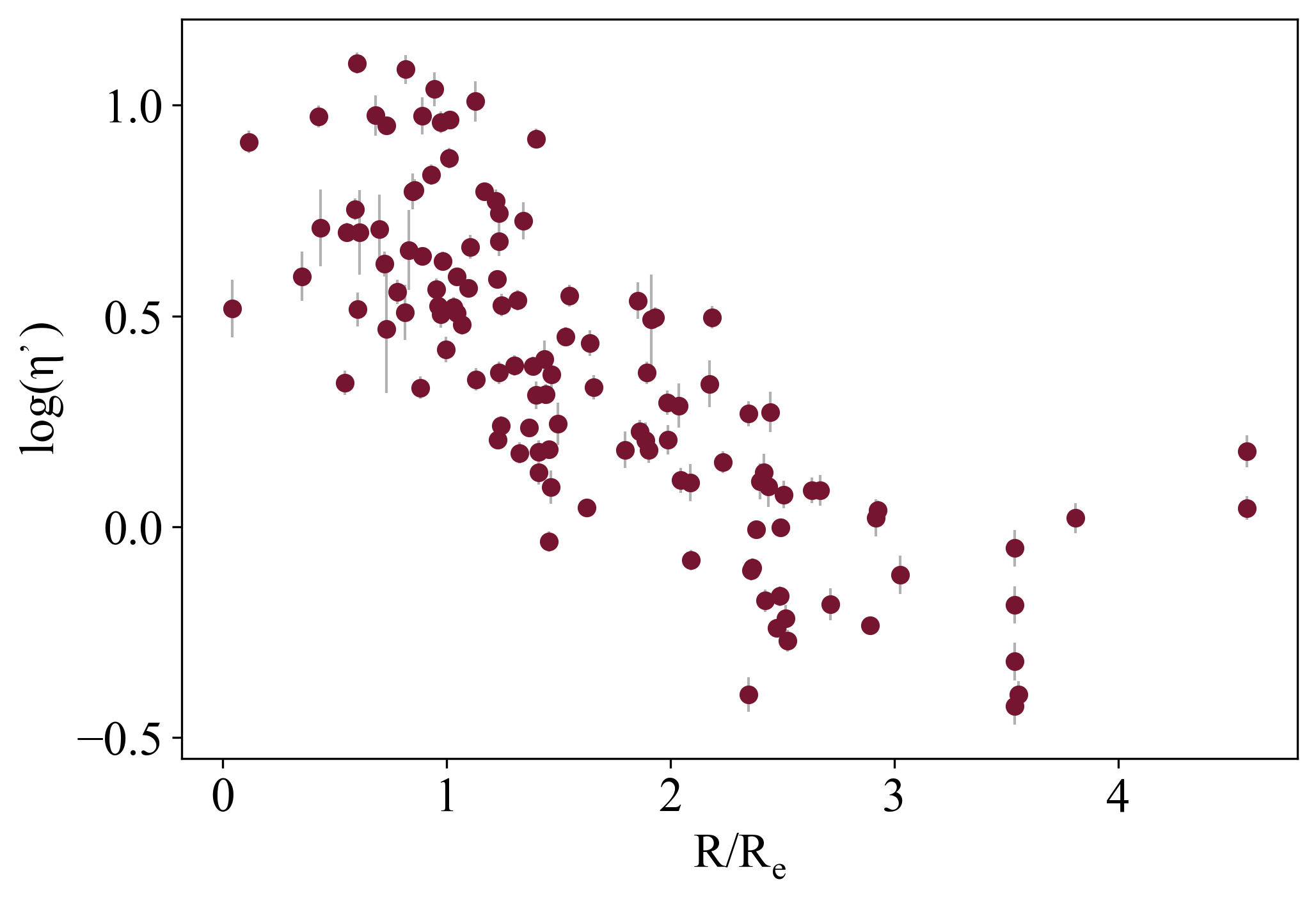}
\includegraphics[width=0.42\textwidth]{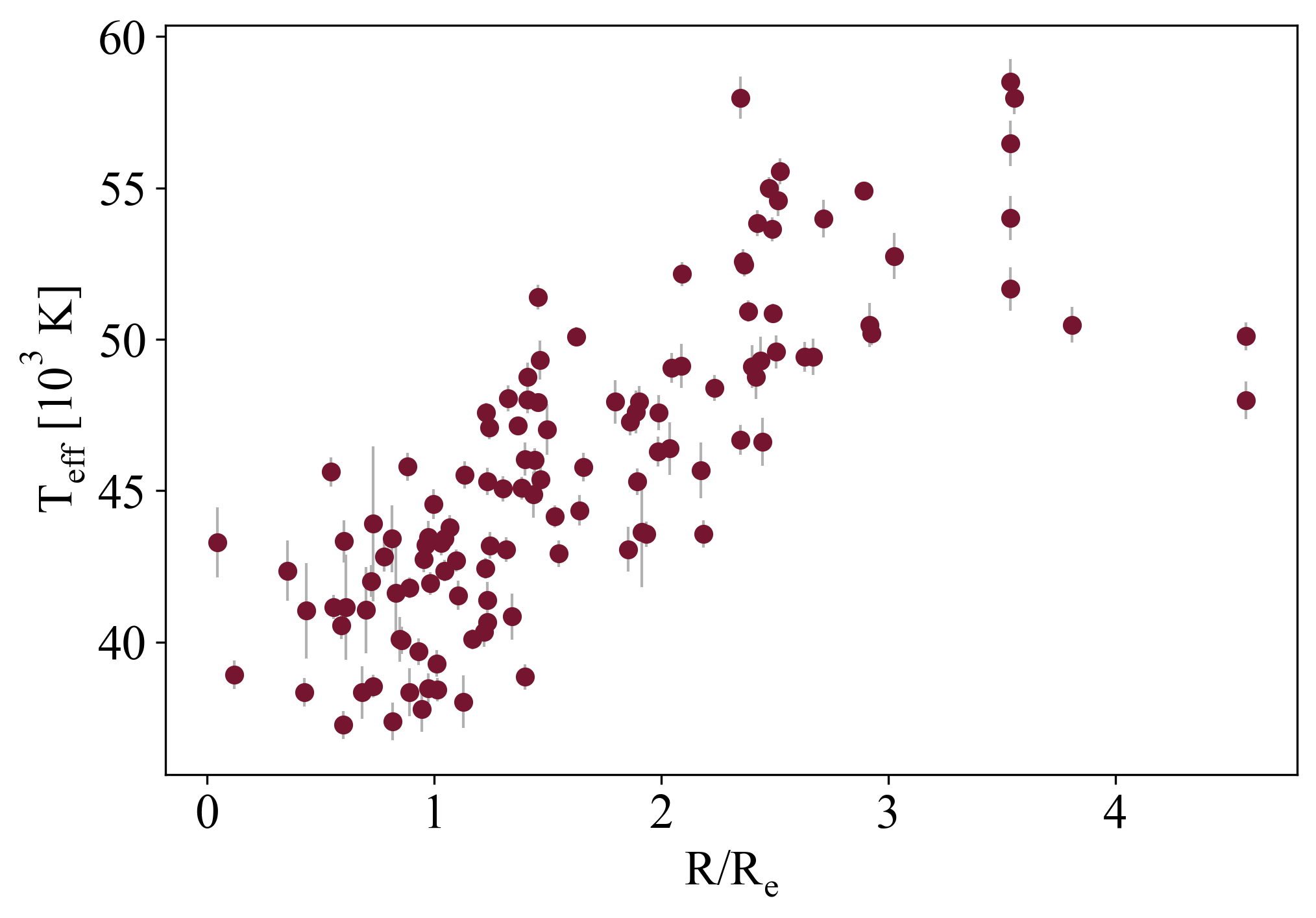}
\caption{Radial gradients of $\eta'$ and the effective temperature of the ionising stellar populations (top and bottom panel respectively).
}
\label{fig_grad_QH_QHe}
\end{figure}

The upper panel of Figure \ref{fig_grad_QH_QHe} shows the radial distribution of $\log(\eta')$, revealing a clear negative gradient, with higher values in the inner regions. This translates into a positive radial gradient in the effective temperature, with a variation of $\Delta\log(\eta')= 0.73$. According to the definition of $\eta'$, higher values correspond to lower effective temperatures of the ionising stars. Therefore, the observed gradient indicates cooler ionising stellar populations towards the galactic centre, consistent with the higher stellar metallicities in the inner regions of M101.

This relation can be quantified using the empirical calibration proposed by \citet{Kennicutt2000} for extragalactic HII regions. The ionising stars' effective temperature gradient in M101 has been investigated by several authors \citep[e.g.][]{Shields1976,Evans1986,Vilchez_pagel1988,Henry1995} with inconclusive results. While \citet{Shields1976} found an increase of this effective temperature from the intermediate to the outermost HII regions, with $\Delta\log(T_{eff})=0.02$--$0.13$, \citet{Dors2003} found no evidence for such a gradient. Nevertheless, using the empirical calibration presented above, we find a positive radial gradient in the effective temperature in our sample (see lower panel of Figure \ref{fig_grad_QH_QHe}), with cooler ionising stars in the inner regions. 

The empirical calibration relating $\eta'$ to the effective temperature of the ionising stars provides a way to estimate their characteristic temperatures in HII regions over a broad temperature range, approximately 33000 K, with temperatures extending up to 50000 K, and is directly linked to the stellar spectral classification system rather than relying on a chain of nebular models. However,  the derived temperatures are subject to an rms uncertainty of $\sim$ 3000 K, and may also be affected by metallicity and ionisation parameter variations. These limitations are particularly important in the inner regions of galaxies, where higher metallicities can significantly affect the effective temperatures derived from this calibration. As shown by \citet{Kennicutt2000}, the effective temperature of ionising stars in Galactic HII regions decreases from $\sim$ 55000 K at $12+\log(\mathrm{O/H})<8.2$ to $\sim$ 40000 K at abundances above solar ($12+\log(\mathrm{O/H})\gtrsim8.9$), with most of this decrease occurring below solar metallicity. This makes it difficult to identify effective temperature gradients using $\eta'$, as also noted by \citet{Bresolin1999}.

\subsection{Ionisation parameter}

The ionisation parameter, log(u), represents the strength of the ionising radiation, described as the ratio between the ionising photon flux  to the density of  free electrons in the nebulae. It is commonly written as $u = Q(H)/4\pi c R^{2}n_{e}$ in therms of the number of hydrogen ionising photons $Q(H)$, the nebular radius, $R$, the electron density, $n_{e}$, and the speed of light, $c$,  as a normalization to keep the parameter dimensionless.

\begin{figure}[t!]
\centering
\includegraphics[width=0.42\textwidth]{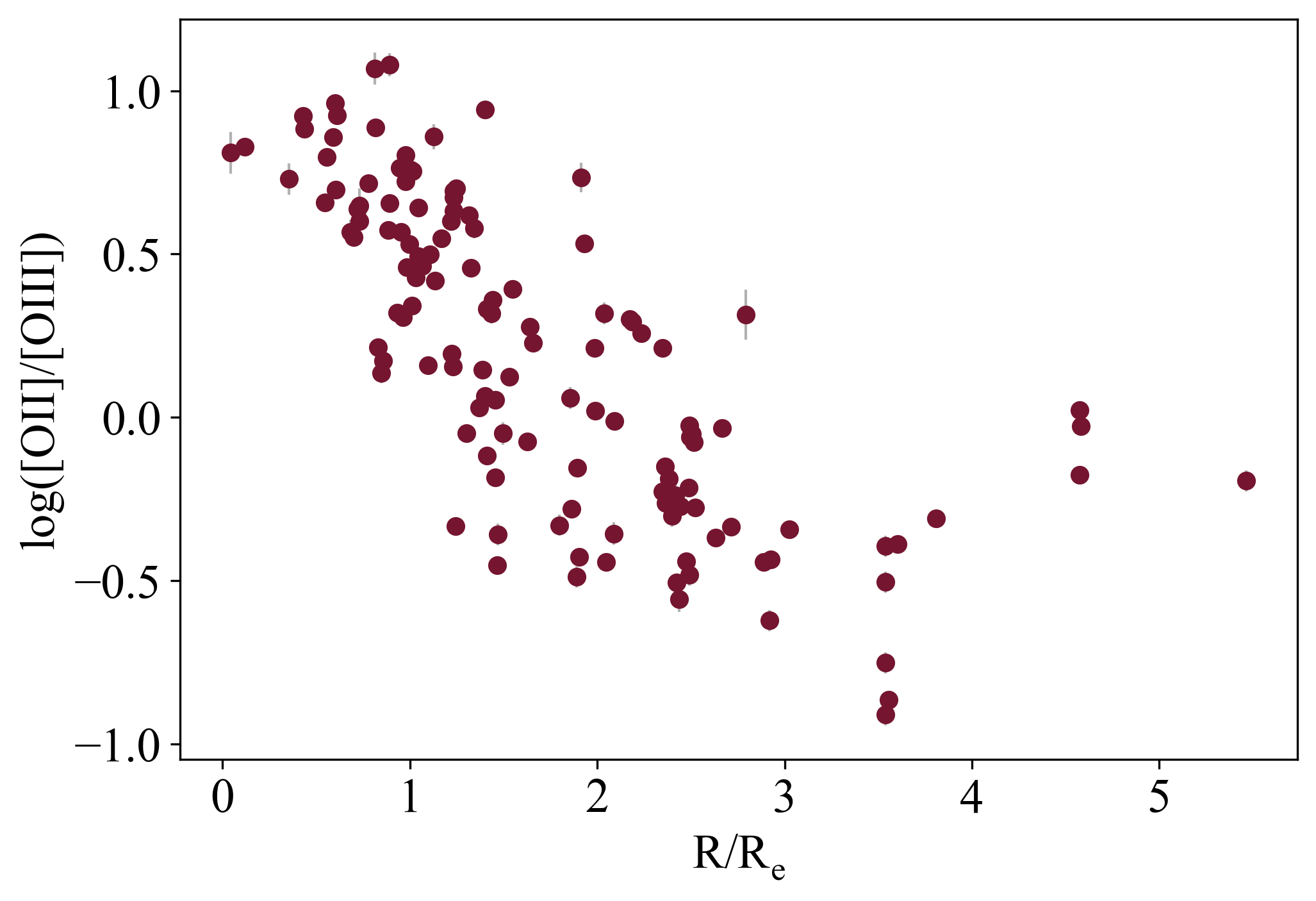}
\includegraphics[width=0.42\textwidth]{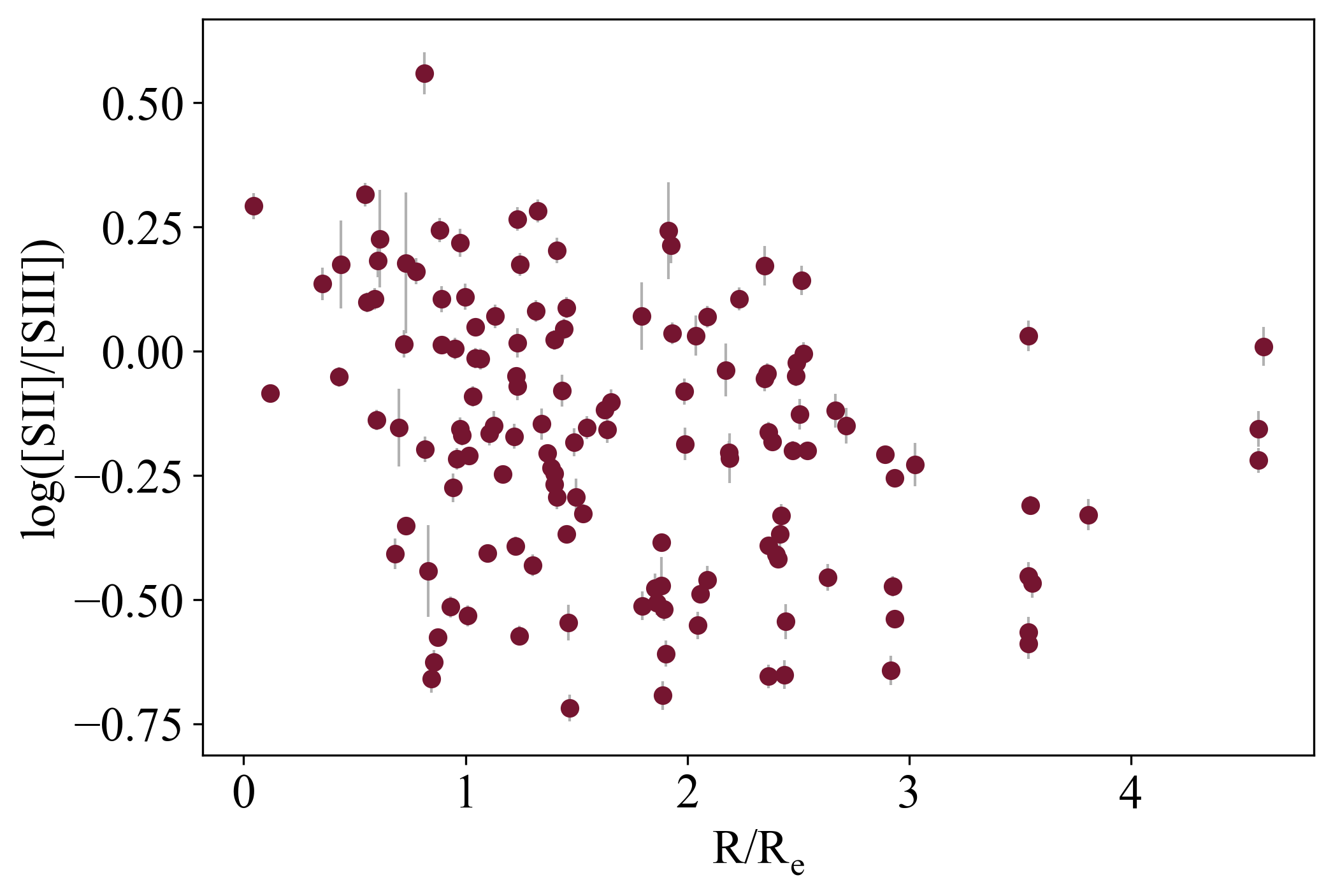}
\includegraphics[width=0.42\textwidth]{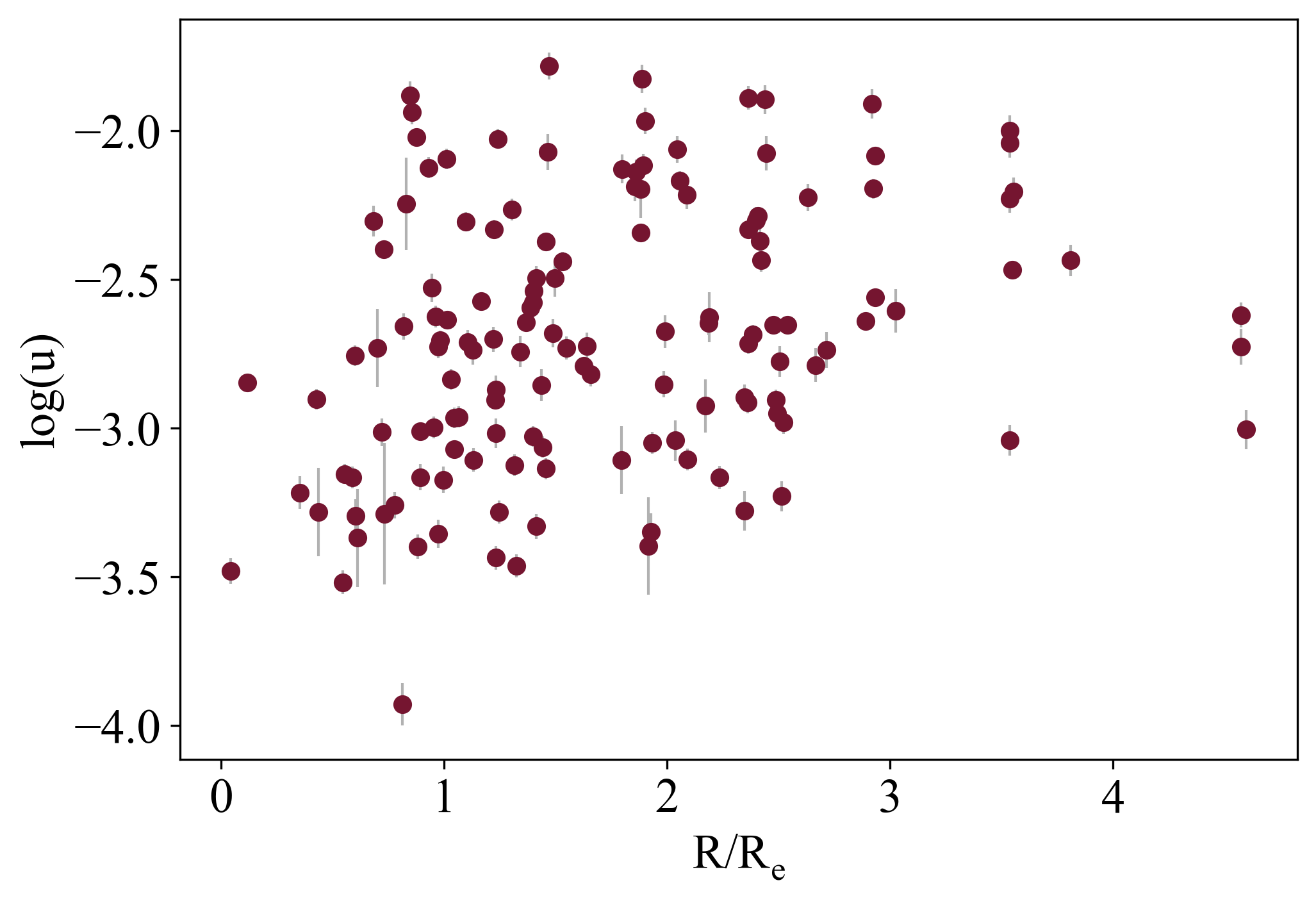}
\caption{Radial gradient of the $[OII]\lambda\lambda3727, 3729$ / $[OIII]\lambda\lambda4959, 5007$ and $[SII]\lambda\lambda6717, 6731$ / $[SIII]\lambda\lambda9069, 9532$ emission-line ratios, and the ionisation parameter derived from the sulphur line ratio following \citet[][]{Diaz2000} (from top bottom panels respectively).}
\label{fig_grad_ionpar}
\end{figure}

The ionisation parameter can be derived from the emission line ratio of two subsequent ionic species of the same element, being the $[OII]\lambda\lambda3727,31 / [OIII]\lambda\lambda4959,5007$ ratio the most widely used option. Adopting the two zone model, we assume a hight electron temperature zone where the high ionisation species are found $O^{2+}$ (35.12 eV). And a low $T_{e}$ zone of low ionisation species like the $O{+}$ ion (13.62 eV). Therefore, as the ionisation  parameter increase, the high electron temperature region grows, since more ionising photons per free electron necessary increase the local equilibrium $T_{e}$ of the gas, reducing the relative size of the low $T_{e}$ zone, and \textit{vice-versa}.
In M101, this  ratio presents a clear trend along the radial profile (see the top panel of Figure \ref{fig_grad_ionpar}), suggesting a lower ionization parameter trend towards the central part of the galaxy. However, upon further inspection, works like \citep{Perez-Montero2005} found another strong correlation for the $[OII]/[OIII]$ ratio with respect to $T_{eff}$ of the ionising source of the nebulae. Since the $O^{+}$ have a high ionization potential (35.12 eV), low $T_{eff}$ ionising sources emit a low number of $O^{+}$ ionising photon ($Q([OII])$). Therefore, the $O^{2+}$ zone of the ionization structure shrinks independently of the ionisation parameter. This dependence on $T_{eff}$ is primarily mediated through a dependence on metallicity, since more metallic O and B stars are dominated by strong abortions in the UV spectral range; explaining the $[OII]/[OIII]$ radial trend.

Nonetheless, the same rational described for oxygen apply to the $[SII]\lambda\lambda6717,31 / [SIII]\lambda\lambda9069,9532$ ratio. With the advantage that the $S^{+}$ ionization potential is lower (23.34 eV) than the oxygen equivalent. Thus the $S^{2+}$ ion dominate the intermediate electron temperature region of the nebulae, and it is less affected by $T_{eff}$ variations. Therefore, we used the \citet{dia91} relation to derived the ionization parameter which can be seen in Figure \ref{fig_grad_ionpar}. Although this value increase towards the galaxy centre, we do not find any clear correlation between $log(u)$ and the radial position within the galaxy.

As a final note, the $[OII]/[OIII]$ and $[SII]/[SIII]$ rate combinations defines the $\eta'$ parameter, which closely follows the $T_{eff}$ of the ionising source, and can be explain by the fact that the log(u) dependence of both ionic ratios cancel out.

\subsection{Oxygen abundance}\label{sec_abundances_O}

Most of the oxygen in ionised nebulae is in the form of $O^+$ and $O^{2+}$, with their relative contributions depending on the degree of ionisation. Only in exceptionally high-excitation objects, typically those showing $HeII$ $\lambda4686$ emission, does $O^{3+}$ make a non-negligible contribution. In our case, this contribution can be neglected. The $O^+/H^+$ and $O^{2+}/H^+$ ionic abundance ratios are derived using $T_e[OII]$ and $T_e[OIII]$, respectively, following the expressions given by \citet{Diaz_Zamora2022}. The total oxygen abundance is then calculated as $O/H = O^+/H^+ + O^{2+}/H^+$.

\begin{figure}[t!]
\centering
\includegraphics[width=0.42\textwidth]{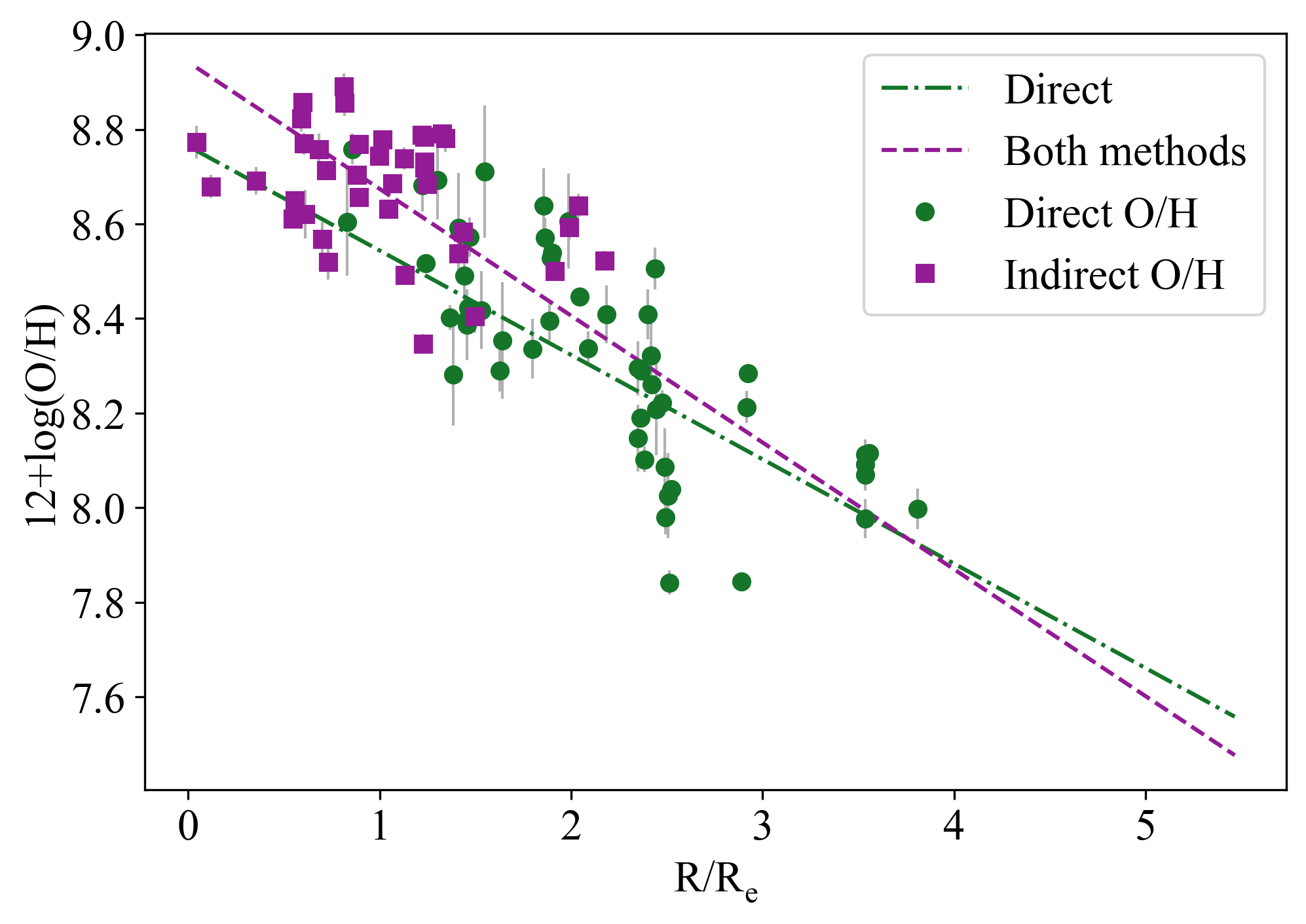}
\includegraphics[width=0.42\textwidth]{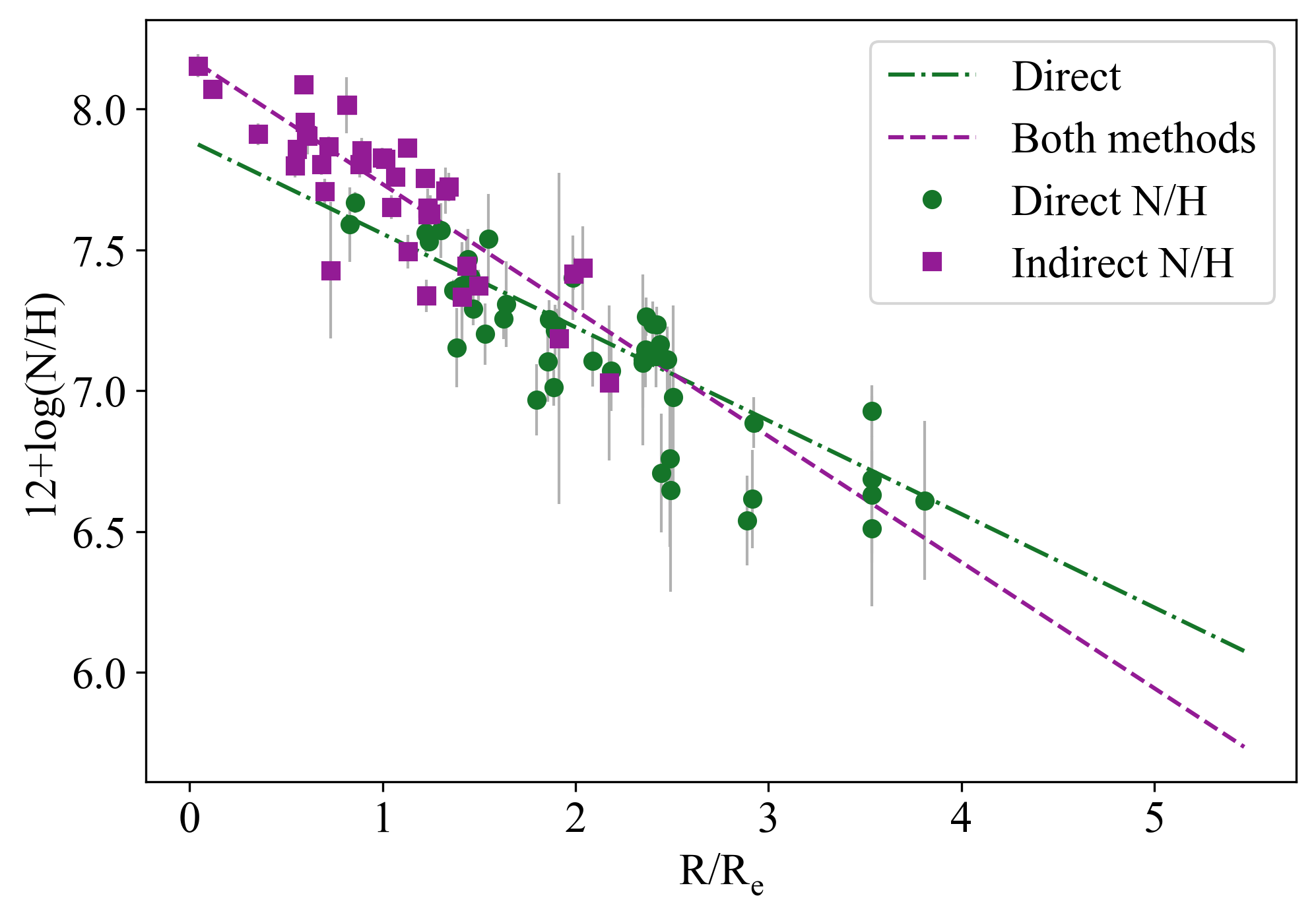}
\caption{Radial gradients of oxygen and nitrogen abundances (top and bottom respectively), derived using the direct (auroral line) and indirect methods. Green circles and purple triangles represent direct and indirect abundance determinations, respectively. Linear fits to the direct method measurements only and to the combined sample from both methods are shown by the green and purple lines, respectively.}
\label{fig_grad_elementos}
\end{figure}

\input{Tables/Results_gradients}

The upper panel of Figure \ref{fig_grad_elementos} shows the oxygen abundance gradient, distinguishing between the values obtained using the direct and indirect methods. We performed linear fits to derive the abundance gradient, first using only the direct measurements and then combining the direct and indirect estimates, to quantify the impact of the selection effects introduced by requiring the detection of the auroral line. The resulting relations are reported in Table \ref{tab:results_gradients}. When only regions with detected auroral lines are considered, the derived gradient is less steep. Including the indirect estimates allows us to extend the abundance gradient to the inner regions, at $R/R_e \lesssim 1.5$, where direct measurements are unavailable. As a result, the inferred central metallicity increases from 8.76 to 8.94, corresponding to an increase of 0.6 dex in solar metallicity units, or a factor of 1.5. At the effective radius, the metallicity difference between the two profiles is 0.13 dex.

\subsection{Nitrogen abundance}

The ionic abundance of $N^+$ relative to $H^+$ was derived from the intensities of the $[NII]\lambda\lambda6548,6584$ emission lines. We first derived the ionic abundance ratio $N^+/O^+$ following \citet{Hagele2008}. According to the ionisation structure adopted in this work ($T_e[NII]=T_e[SIII]$; see Section \ref{sec_ion_structure}), we refer to abundances derived using directly measured $T_e[SIII]$ values as direct abundances, while those obtained using the assumed temperature relation are referred to as indirect abundances.

The N/O abundance ratio was derived from the $[NII]\lambda\lambda6548,6584$ to $[OII]\lambda\lambda3727,3729$ emission-line ratio (see Figure \ref{fig_NIIOII}). The resulting radial gradient follows the expected behaviour: $N^+/O^+$ remains approximately constant across the outer regions, reflecting primary nitrogen production, and increases towards the galaxy centre as the contribution from secondary nitrogen production becomes more significant.

To convert the ionic abundance ratio into the elemental N/O ratio, we assumed that $N^+$ and $O^+$ have similar ionisation structures in HII regions, such that $N/N^+ = O/O^+$. This assumption allows us to estimate N/O directly from $N^+/O^+$, without applying an additional ionisation correction factor (ICF). The total nitrogen abundance was then obtained by combining the $N/O$ and $O/H$ abundance ratios: $log(N/H) = log(N/O)+log(O/H)$.

The bottom panel of Figure \ref{fig_grad_elementos} shows the nitrogen abundance gradient. Including the measurements obtained with the indirect method results in a steeper abundance gradient and increases the nitrogen abundance at the effective radius by 0.18 dex.

\section{Discussion} \label{discusion}

\subsection{Nitrogen over oxygen gradient}\label{sec_rel_gradients}

\begin{figure}[t!]
\centering
\includegraphics[width=0.42\textwidth]{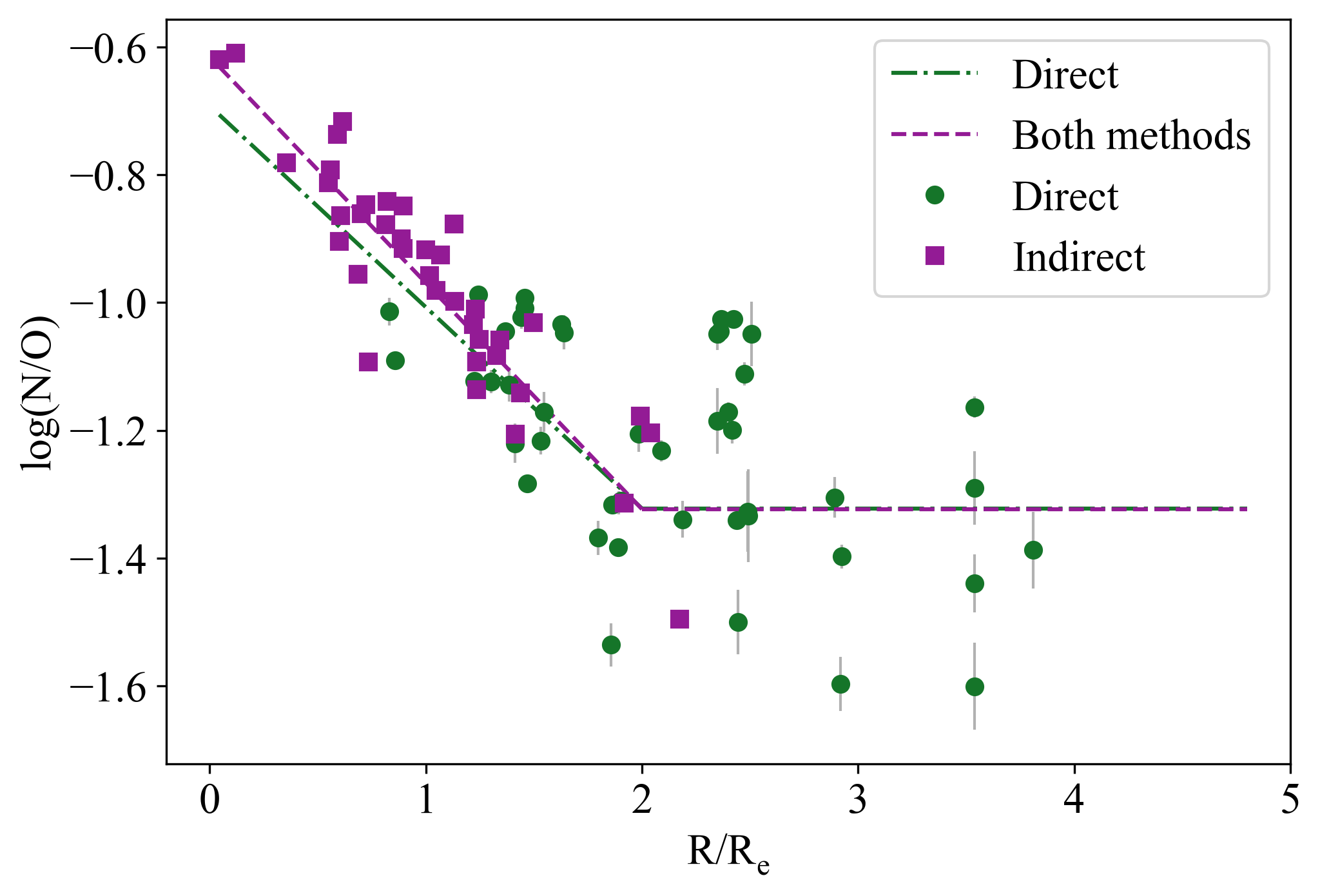}
\caption{Radial abundance gradients of log(N/O). Green circles and purple triangles represent direct and indirect abundance determinations, respectively. The linear fits to the direct measurements only and to the combined sample of direct and indirect measurements are shown by the green and purple lines, respectively, and they are restricted to the inner regions of the gradient ($R<2R_e$).} \label{fig_grad_elementos_relativos}
\end{figure}

Nitrogen is produced by the CN-branch of the CNO cycle in the H burning zone of the stellar interior. This reaction takes place mainly in intermediate mass stars from 4 to 8 $M_{\odot}$, that release their metal content throughout the asymptotic giant branch (AGB) evolutionary phase with a typical time delay of $\sim$ 250 Myr respect to the oxygen production \citep{Henry2000}. Nitrogen nucleosynthesis is controlled by the presence of carbon in the stellar core and therefore presents the so called primary and secondary relation to the initial metal content of the star. Primary production is fixed by the star own synthesis of carbon and nitrogen, while the secondary nature refers to the linear increase in N production derived from C already present in the star at birth time. Furthermore, the initial carbon seed in the stellar core is not the only effect of metallicity in N production, since the increased metal content of the star modifies the stellar wind driven mass loss rate, affecting the late phases of stellar evolution \citep{Meynet2005}.

As can be seen in Figure \ref{fig_grad_elementos_relativos}, our N/O radial distribution reproduces the expected behavior of a constant plateau dominated by dispersion in the outer part of the gradient, and a linear increase trend towards the central part of the galaxy. As reported in previous works \citep{PerezMont2009,Vincenzo2016,Berg2020,Esteban2020,ArellanoC2020,Roy2021,Rogers2022}, the dispersion in the N/O plateau comes from the effects of the particular SFH of the regions, the cumulative effects of pristine gas inflows and the local nitrogen pollution by Wolf-Rayet stars. While the low scatter trend shown in the inner HII regions, the $R< 2 R_{e}$ part of the galaxy is explained by the secondary production of nitrogen. It is worth mentioning the importance of the indirect abundance determination method to study N/O relative abundances in the inner part of the galaxy, where auroral line detection is almost unfeasible.

\subsection{Comparison with previous works}

The first comparison with previous studies concerns the electron temperature derived from auroral line measurements reported by different authors and obtained for independent observations. Figure \ref{fig_te} shows there is a significant scatter in the effective temperature values among regions located at similar galactocentric radii. A closer inspection reveals that at $R/R_e \sim 2.5$, we have measurements of the GEHR NGC 5447 from both \citet{Croxall2016} and \citet{Kennicutt2003}, as well as an additional SDSS observation (SDSS-14). This region is resolved into four distinct knots: H143 (NGC5457-371.1-280.0), H149 (NGC5457-368.3-285.6), H128 (NGC5457-392.0-270.1) and H132 (SDSS-14 in our sample). Interestingly, all the $T_e[SIII]$ measurements reported by \citet{Kennicutt2003} for these knots are below $10000 K$, whereas those reported by \citet{Croxall2016} show systematically higher values, reaching approximately $14000 K$. Another region located at a similar galactocentric radius is H71 (NGC5457-453.8-191.8) right next to H77 (SDSS-16 in our sample), for which two independent measurements are available and show very similar $T_e[SIII]$ values ($\sim 14000 K$). A similar situation is found at $R/R_e \sim 3.5$ actually correspond to only two regions, each comprising several distinct knots: NGC5457+667.9+174.1 (SDSS-19) and N5471 (knots A, B, C, and D). This highlights the importance of carrying out a detailed analysis of individual knots and their local properties within the same HII region.

We also compared our abundance gradients with those reported by \citet{ken03}, \citet{Croxall2016}, \citet{Esteban2020}, and \citet{Berg2020}, which are based on samples of different sizes and radial coverage. \citet{ken03} derived electron temperatures and chemical abundances for 20 HII regions in M101, spanning a radial range of 2.74–18.03 arcmin. \citet{Croxall2016} studied 109 HII regions, obtaining robust measurements of one or more temperature-sensitive auroral lines for 74 of them, over a radial range of 0.39–15.08 arcmin. The sample analysed by \citet{Esteban2020} comprises 11 HII regions with direct electron-temperature measurements, covering 6.20–18.03 arcmin. Finally, \citet{Berg2020} obtained direct electron-temperature measurements for 190 individual HII regions, spanning 0.03–2.71 arcmin. However, when restricting the sample to regions with a detection of either the [OIII] or [SIII] auroral lines, the radial coverage is reduced to 1.33–2.71 arcmin. The indirect method proposed in this work allows us to take advantage of all the measurements available in the literature, including those from the innermost regions, where auroral lines are not detected, therefore extending the radial coverage to 0.14–15.14 arcmin.

Figure \ref{fig_grad_literature} compares the abundance gradients reported in previous studies with those derived in this work. For oxygen (top panel), all the gradients based on direct measurements show very similar results. This is likely due to the limitations of relying exclusively on regions where the auroral line is detected, as the [OIII] auroral line is not observable at radii $R\lesssim 1 R_e$ in M101. We also find that the gradient derived using our direct measurements is consistent with those reported in previous studies. However, by applying our new indirect method, we are able to extend the analysis to the innermost regions, resulting in a steeper abundance gradient. 
Additionally, the behaviour of the $N/H$ gradient (bottom panel) in the inner regions, where secondary nitrogen production becomes dominant, is similar to that observed for oxygen, with our fit yielding a steeper slope than the other ones presented. However, in all cases, the derived slopes are strongly influenced by the adopted normalisation of the gradient, as the outer regions are dominated by a plateau with considerable scatter. This dispersion may reflect the effects of the particular star formation histories of individual regions, the cumulative effects of pristine gas inflows, and local nitrogen enrichment by Wolf-Rayet stars (see Section \ref{sec_rel_gradients}).

\subsection{Oxygen abundance and different calibrations}

The direct method has traditionally been used to derive chemical abundance gradients in galaxies. However, several empirical calibrations have also been proposed to estimate abundances directly from ratios of strong emission lines, providing an alternative when auroral lines are too faint to be detected. Here, we use our new method to assess the performance of the main empirical calibrations based on oxygen, sulphur, and nitrogen emission lines.

The $R_{23}$ parameter, first introduced by \citet{Pagel1979}, is one of the most widely used strong line indicators for estimating the oxygen abundance in HII regions and galaxies. It is defined as the ratio of the combined fluxes of the $[OII]\lambda\lambda3727,3729$ and $[OIII]\lambda\lambda4959,5007$ emission lines normalised to the  H$\beta$ flux. It can be used to estimate oxygen abundances across both the metal-poor ($12+log(O/H) < 8.5$) and metal-rich ($12+log(O/H) > 8.5$) regimes. However, its double-valued dependence on metallicity gives rise to a well known degeneracy between the lower- and upper-metallicity branches. This parameter has been extensively studied and calibrated using both theoretical and empirical approaches \citep[i.e.][]{Yin2007,Edmunds1984,Skillman1989,McGaugh1991,Kobulnicky1999,Tremonti2004,Jiang2019}.

Numerous calibrations of the $R_{23}$ parameter have been proposed in the literature over the years. Given the range of oxygen abundances found in our M101 sample (see Section \ref{sec_abundances_O}), we focus here on calibrations that describe the upper-metallicity branch. In particular, we consider three of the most widely used calibrations in the literature for the local Universe: \citet{Nagao2006,Perez-montero2017,Pilyugin2005}.

\begin{figure}[t!]
\centering
\includegraphics[width=0.42\textwidth]{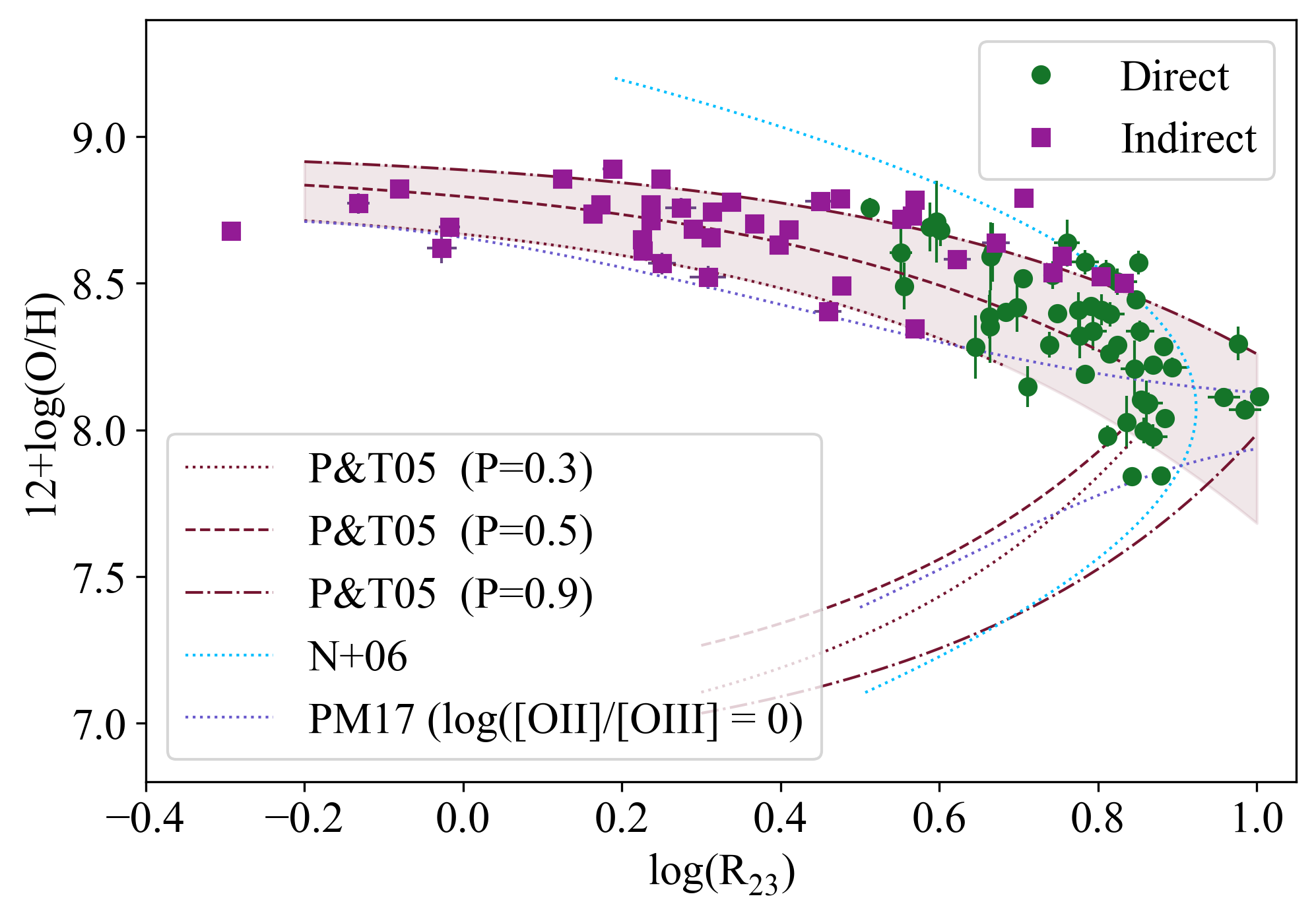}
\includegraphics[width=0.42\textwidth]{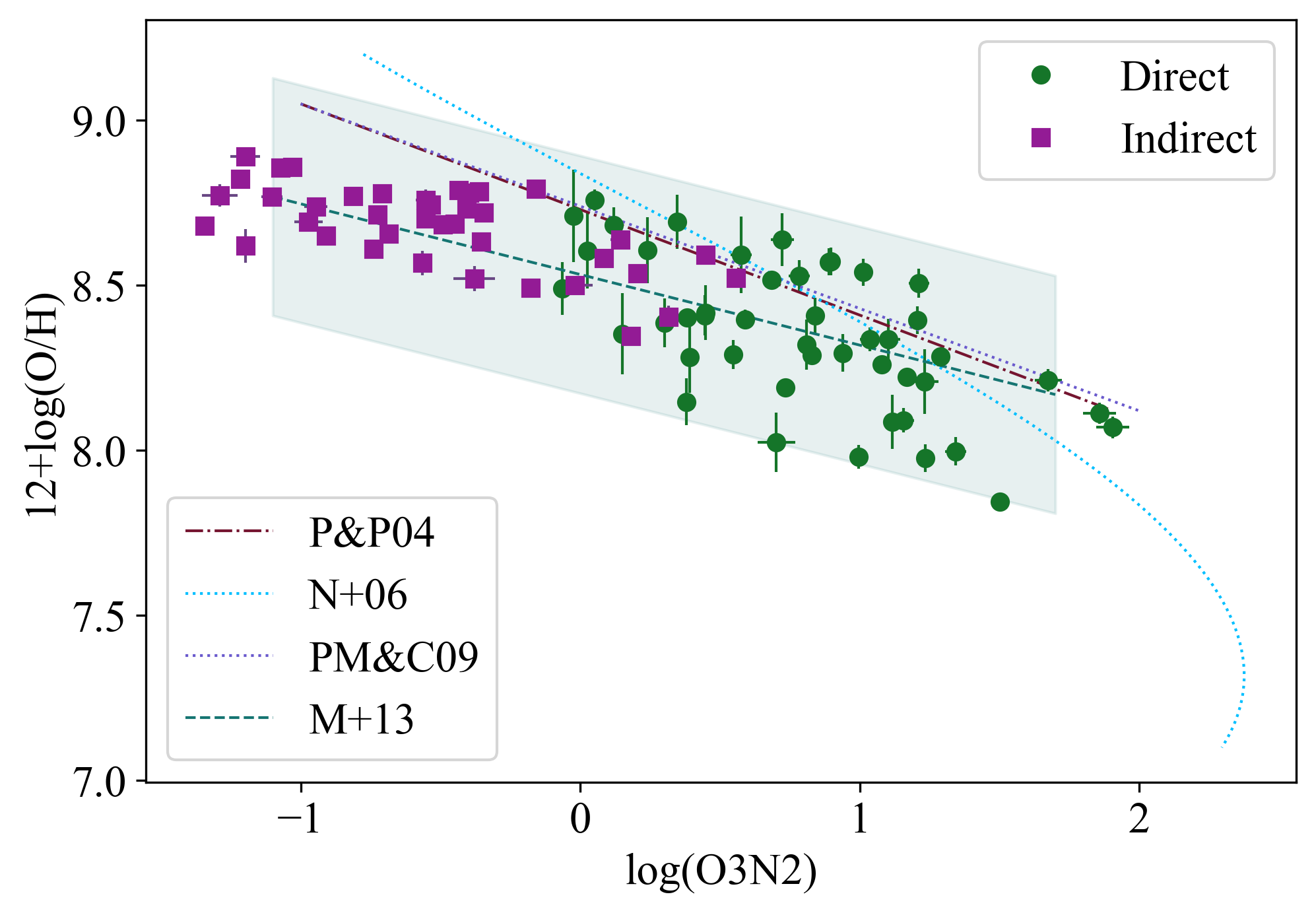}
\includegraphics[width=0.42\textwidth]{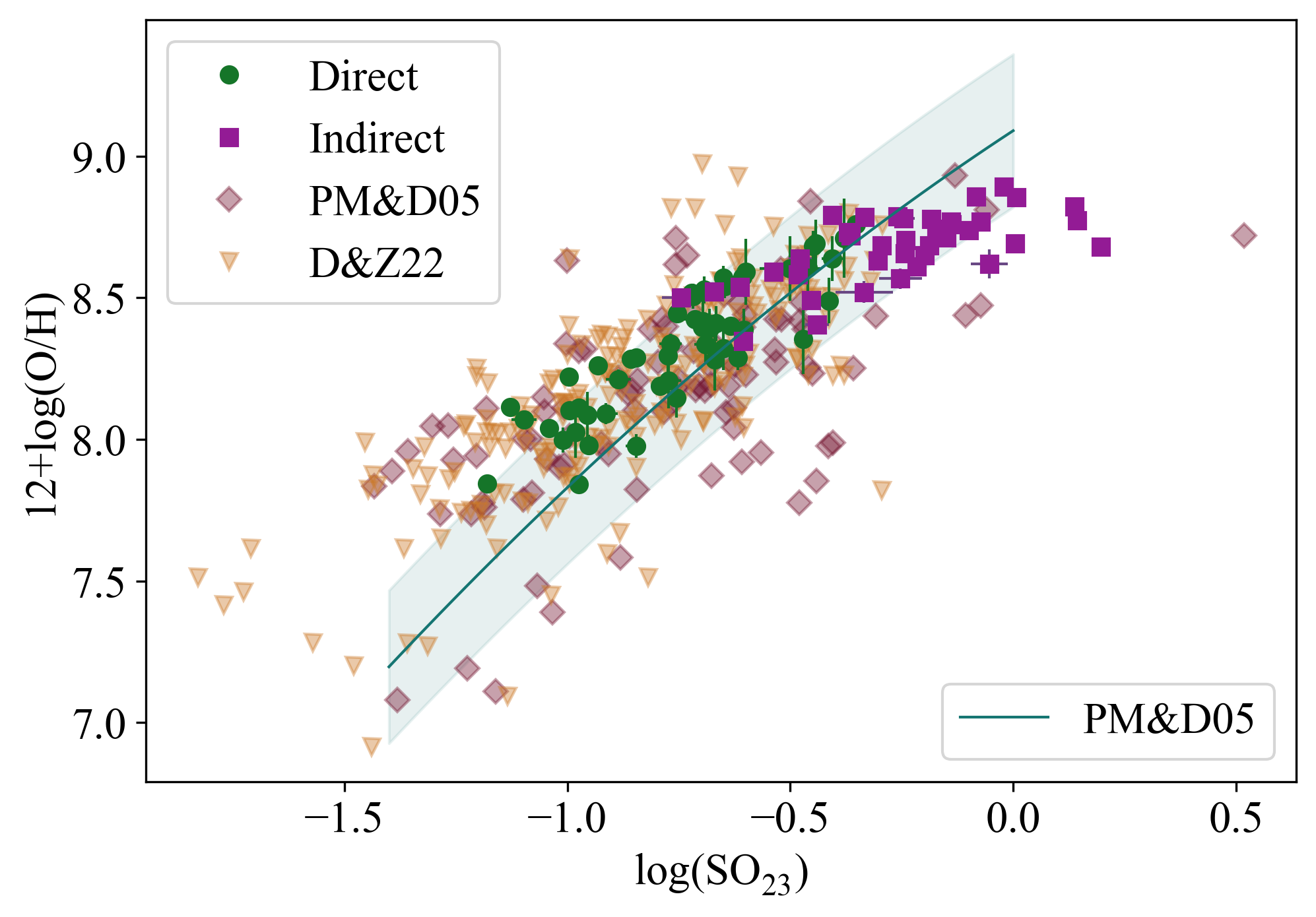}
\caption{Top panel: Comparison of the $R_{23}$ calibrations of \citet[][P\&T05]{Pilyugin2005} , \citet[][N+06]{Nagao2006}, and \citet[][PM17]{Perez-montero2017} with our measurements. The shaded area shows the range covered by P\&T05 after applying the $P$-excitation correction using the values measured in this work. Middle panel: Comparison of the $O3N2$ calibrations of \citet[][P\&P04]{Pettini2004}, \citet[][N+06]{Nagao2006}, \citet[][M\&C09]{PerezMont2009}, and \citet[][M+13]{Marino2013} with our measurements. The shaded area represents the 95\% dispersion range of M+13. Bottom panel: The $SO_{23}$ calibrator of \citet[][PM\&D00]{Perez-montero2000}, applied directly as an abundance proxy, compared with our measurements. The shaded area represents its $\pm 0.27$ dex dispersion. The compilations of D\&Z22 and PM\&D00 are also shown as yellow triangles and red diamonds, respectively. Green circles and purple triangles represent direct and indirect abundance determinations, respectively.}
\label{fig_calibrators}
\end{figure}

The upper panel of Figure \ref{fig_calibrators} shows the $R_{23}$ parameter, together with the corresponding calibrations provided by their authors. Most of the direct abundance measurements obtained in this work, for which the [OIII] auroral line is detected and therefore the $T_e[OIII]$ can be determined directly, lie in the turnover region where the $R_{23}$ relation transitions between the lower- and upper-metallicity branches. This makes it difficult to distinguish between the two branches and introduces significant uncertainties in the abundance estimates derived from the $R_{23}$ calibrations. The small number of direct abundance measurements on the upper branch is one of the main reasons why calibrating this part of the relation remains challenging. However, by including the indirect abundance measurements obtained using the method proposed in this work, we are able to sample almost the entire upper branch. This provides a much stronger basis for constraining the calibrations in the high metallicity regime and highlights the importance of our approach for extending abundance determinations to regions where direct measurements are limited, as discussed above.

The calibration that best reproduces our data is that of \citet{Pilyugin2005}, which accounts for the excitation degree through the parameter \(P\), defined as: $P = [OIII]\lambda\lambda4959,5007$ / $([OII]\lambda\lambda3727,3729$ + $[OIII]\lambda\lambda4959,5007)$. This method defines the calibration coefficients as a function of metallicity, based on the $R_3$–P diagram, with $R_3 = [OIII]\lambda\lambda4959,5007/ H\beta$. Although our measurements fall within the range covered by the calibration proposed by \citet{Pilyugin2005} (see Figure \ref{fig_calibrators}), the excitation correction, P, can lead to substantially different metallicity estimates for a given $R_{23}$. This effect is particularly relevant for the upper-metallicity branch, where the R$_{23}$ relation becomes relatively flat. In this regime, a given value of P can correspond to a broad range of $R_{23}$ values, making the derived abundance particularly sensitive to the excitation parameter. Since $P$ is defined in terms of emission-line intensities, it is also sensitive to changes in the effective temperature of the ionising stellar population ($T_{eff}$) and the ionisation parameter (log($u$)). Both quantities can vary significantly along a metallicity gradient, as discussed in Section \ref{sec:res}, potentially introducing additional uncertainties into abundance estimates on the upper branch. This discrepancy may be related to uncertainties in the adopted emissivities, which were not apparent in the original calibration based on only four regions, and whose authors already cautioned against its use at high metallicities due to the lack of observational constraints.

The top panel of Figure \ref{fig_piliyuigin} in Appendix \ref{ap:fig}, shows this diagram, with the different relations corresponding to the metallicity range covered by our sample ($12+log(O/H) \sim 7.6 - 8.8$) and the coefficients reported by the authors. Our results show no clear dependence on metallicity in the way in which it is parametrised in their calibration. In fact, a single relation is sufficient to reproduce the measurements across the entire abundance gradient. This behaviour is also illustrated in the bottom panel of the same figure, which shows the relation between P and oxygen abundance. Neither of the two samples considered separately shows any clear correlation. Although the combined sample appears to follow a general trend in which P decreases with increasing $O/H$, the large scatter around this relation implies substantial uncertainties in the corresponding abundance estimates. Furthermore, since lower values of P are preferentially found at higher metallicities, this dependence could introduce an artificial flattening of the metallicity gradient.

\begin{figure}[t!]
\centering
\includegraphics[width=0.42\textwidth]{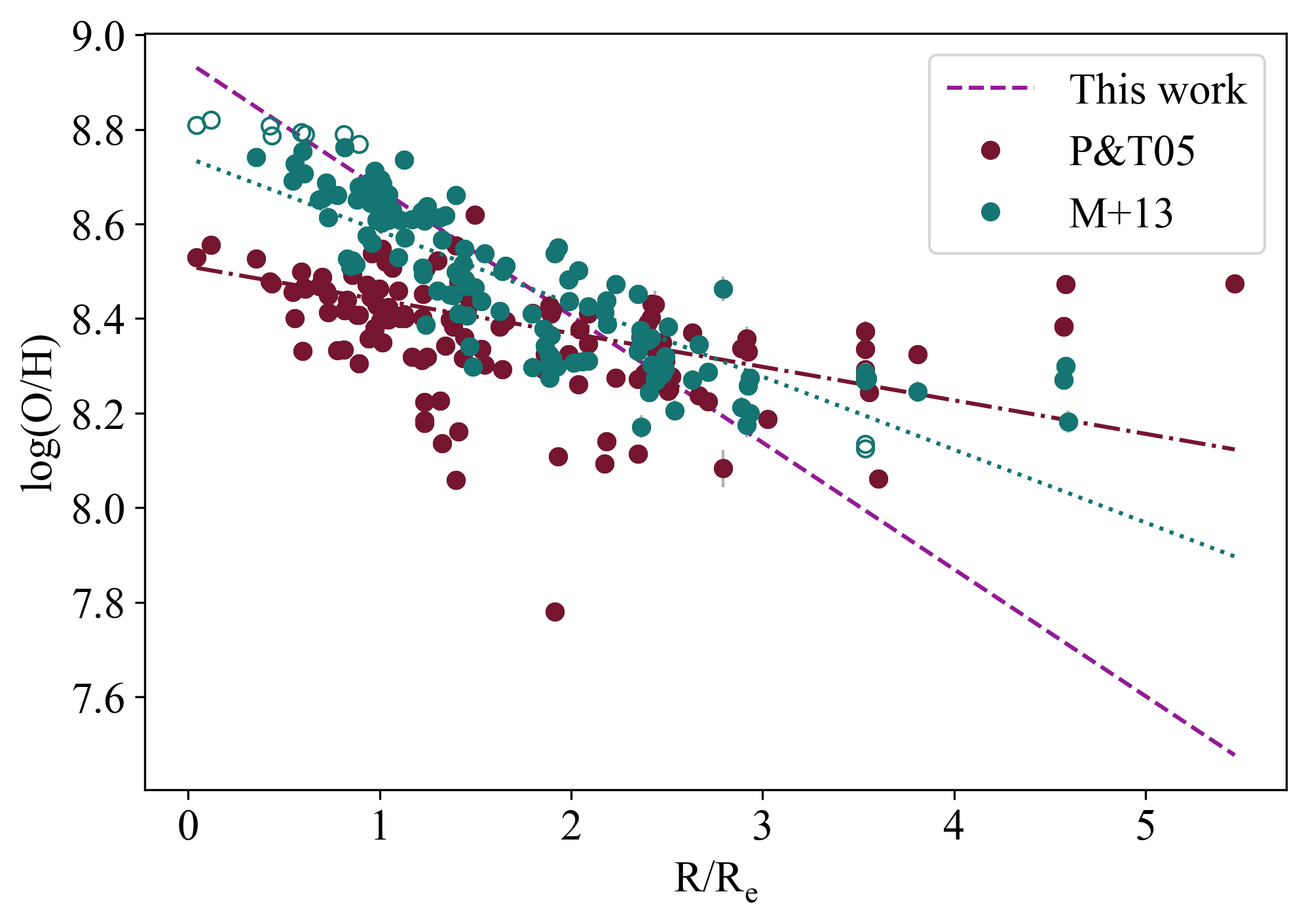} 
\caption{Oxygen abundance gradients derived using the $R_{23}$ and $O3N2$ calibrations proposed by \citet{Pilyugin2005} and \citet{Marino2013}, shown as red and blue circles, respectively. The corresponding radial fits are plotted in the same colours. Open circles indicate measurements outside the validity range of the corresponding calibrations.}
\label{fig_grad_indirect}
\end{figure}

Figure \ref{fig_grad_indirect} shows the oxygen abundance gradient derived using the \citet{Pilyugin2005} method with our measurements. The effect described above is clearly visible: the dependence on P artificially flattens the gradient for the inner regions ($R < 1 R_e$), while the uncertainties in the turnover of the calibration overestimate the oxygen abundance for the outer regions ($R > 3 R_e$).

Another widely used strong-line indicator for estimating oxygen abundances is $O3N2$, first introduced by \citet{Alloin1979} and defined as the ratio of the $[OIII]\lambda5007$/H$\beta$ and $[NII]\lambda6583$/H$\alpha$ emission-line ratios. Compared with $R_{23}$, $O3N2$ is less affected by dust extinction because each forbidden line is normalised to a nearby Balmer line. Moreover, the emission lines involved are at relatively long wavelengths, further reducing the impact of dust attenuation. $O3N2$ relies on some of the strongest and most readily accessible emission lines in rest-frame optical spectra. Its upper-metallicity branch extends down to relatively low abundances ($12+\log(O/H)\gtrsim 7.5$), making it particularly useful for abundance studies over a broad metallicity range and for determining radial abundance gradients \citep[e.g.][]{Sanchez-Menguiano2016}. One of the most widely used calibrations of $O3N2$ was introduced by \citet{Pettini2004}, followed by several subsequent calibrations, including those proposed by \citet[][N+06, PM\&C09, and M+13, respectively]{Nagao2006,PerezMont2009,Marino2013}. 

The middle panel of Figure \ref{fig_calibrators} shows the calibrations discussed above together with our measurements, highlighting the best agreement with the calibration proposed by \citet{Marino2013}. As in the case of $R_{23}$, the indirect method proposed in this work allows us to include a substantially larger number of measurements at high metallicities, extending the range covered by the calibration down to $O3N2$ $\sim -1.5$. However, even after propagating both the measurement errors and the uncertainties associated with the calibration, the abundance estimates remain subject to considerable uncertainties due to its large intrinsic dispersion. According to \citet{Marino2013}, 68\% and 95\% of the measurements fall within $\pm 0.18$ dex and $\pm 0.36$ dex, respectively, highlighting the limitations of this calibration for accurately determining abundances in the high-metallicity regime.

Figure \ref{fig_grad_indirect} shows the oxygen abundance gradient derived using our measurements and the \citet{Marino2013} calibration. Even in M101, a galaxy that is not particularly metal-rich, the validity range of this calibration ($-1.1 < O3N2 < 1.7$) does not extend to the inner regions, which lie beyond its high-metallicity limit. Moreover, although the calibration is formally valid in the outer regions, the derived abundances show a saturation effect due to its turnover.

\subsection{Advantages of calibrating $T_e$ rather than the chemical abundances}
Empirical abundance calibrations have traditionally been developed based on the exponential dependence of collisionally excited line emissivities on electron temperature. As the metallicity increases, metal cooling becomes more efficient, leading to lower electron temperatures and consequently weaker auroral lines, which may become too faint to be detected. Following this approach, numerous empirical calibrations have been developed to estimate metallicities directly from the strong emission line rations, relating their intensities to the chemical abundance \citep[i.e.][]{Jensen1976,Alloin1979,Pagel1979}.

However, other physical parameters also play an important role, as HII regions are not single-parameter systems and they are not governed by metallicity alone. At a given metallicity, they can exhibit different effective temperatures ($T_{eff}$), ionisation parameters ($u$), and other physical conditions (see Section \ref{sec:res}). Our approach calibrates the electron temperature first and then derives the chemical abundance separately, allowing the physical conditions of the ionised gas to be explicitly taken into account. This provides further insight into the ionisation structure of the nebulae, which can then be used to apply the appropriate ionisation corrections and obtain a more reliable estimate of the total chemical abundance.

The $SO_{23}$ parameter used in this work to estimate $T_e[SIII]$ was employed in previous studies to derive oxygen abundances directly, without explicitly determining the ionic abundances. The bottom panel of Figure \ref{fig_calibrators} shows the calibration proposed by \citet{Perez-Montero2005}, together with the literature compilation used in that work, the compilation by \citet{Diaz_Zamora2022}, and the direct and indirect abundance measurements obtained in this study. The relation remains single-valued across the entire metallicity range, avoiding the complications associated with the two branches and the turnover region. The limited number of high abundance measurements available when the original calibration was developed is reflected in its upper end, which fails to reproduce both the rightmost data point and the new indirect abundance measurements obtained for M101. At lower metallicities, the inclusion of additional measurements also reveals a considerably larger dispersion than that originally reported by \citet[][$\pm$ 0.27 dex]{Perez-Montero2005}. This larger scatter may be related to the sensitivity of the calibration to the physical conditions of the nebulae, particularly their ionisation degree and effective temperature. Indeed, $SO_{23}$ = $S_{23}$/$O_{23}$ and it retains the same sources of uncertainty as its two parent indicators, since the $S_{23}/O_{23}$ ratio increases at lower ionisation degrees and lower effective temperatures \citep{Perez-Montero2005}.

Nevertheless, these limitations do not apply in the same way when the calibrator is used to estimate $T_e[SIII]$. Since our new indirect method is based on the cooling of the nebula, the temperature determination does not require explicit corrections for its ionisation structure or other physical conditions. In contrast, when $SO_{23}$ is used to estimate oxygen abundances directly, the different contributions from $O^+$ and $O^{++}$ must be taken into account. The relative abundances of these ions vary with the ionisation conditions: in low-metallicity regions, oxygen is predominantly in the form of $O^{++}$, whereas in more metal-rich regions, $O^+$ becomes increasingly important \citep[see, e.g.][]{Berg2020,Diaz_Zamora2022}. This variation in the ionic balance, combined with the lower electron temperatures characteristic of metal-rich regions, introduces additional complexity into the abundance calibration. 

\subsection{New ionisation structure and advantages at high abundances}

The ionisation structure commonly adopted in the literature for the determination of ionic abundances assumes two distinct zones. In the high ionisation zone, $O^{++}$ is the dominant oxygen ion, and the electron temperature is assumed to be well represented by $T_e[OIII]$. In the low ionisation zone, $O^+$ is the dominant ion, and the electron temperature is generally assumed to be represented by $T_e[OII]$. Since $T_e[OII]$ cannot usually be measured directly, it is commonly estimated using relations derived from photoionisation models \citep[e.g][]{Stasinska1990,Pagel1992,Izotov1994}.

The first limitation of this commonly adopted ionisation structure is that the relation between $T_e[OII]$ and $T_e[OIII]$ is not straightforward, and no single relation between the two temperatures has been established. In particular, $T_e[OII]$ also depends on the electron density, as shown by \citet[][see their Fig. 8]{Hagele2006}. A considerable scatter is also observed in relatively homogeneous samples of objects, even when accounting for the measurement uncertainties. As illustrated by \citet[][see their Fig. 4]{Izotov2006}, most of the measurements with smaller error bars do not follow the commonly adopted relation between the two temperatures.

The second limitation of the commonly adopted ionisation structure arises from the temperature sensitivity of collisionally excited lines. Since emissivities increase exponentially with electron temperature, the observed line intensities are strongly weighted towards warmer parts of the nebular ionisation structure \citep{Shields1978}. Consequently, the temperature inferred from the [OIII] line ratios may not be representative of the actual thermal conditions of the $O^{++}$ emitting gas, leading to an overestimation of its ionic temperature, such that $T_e[OIII]>T(O^{++})$. Since the temperature of the $O^+$ is commonly estimated from that of $O^{++}$, this bias is also propagated to the low ionisation zone, resulting in an underestimation of the $O^+$ abundance \citep[see also][]{Zamora2026}.

The impact of these two biases on the determination of the total oxygen abundance depends on the metallicity of the nebulae. In low metallicity HII regions, where most of the oxygen is in the form of $O^{++}$, uncertainties in the adopted value of $T_e[OII]$, can lead to significant uncertainties in the $O^+/H^+$ abundance ratio. However, since $O^{++}$ dominates the oxygen budget, their impact on the total $O/H$ abundance remains relatively small. The situation is reversed in metal-rich regions, where $O^+$ becomes the dominant oxygen ion and uncertainties in $T_e[OII]$ can therefore propagate into substantial uncertainties in the total oxygen abundance. In fact, this effect leads to an underestimation of the $O^+/H^+$ abundance, which is the main contributor to the total $O/H$ abundance. It is important to emphasise that this effect is already relevant at relatively moderate, non-extreme metallicities, such as those typically found in disc galaxies. As shown by \citet[][see their Fig. 15]{Diaz_Zamora2022}, this effect is also present in regions with typical median values of $12+\log(S/H) \sim 6.92+1.57$ and $T_e[SIII] \sim 7900$ K.

We mitigated both biases by deriving $T_e[OII]$ independently of $T_e[OIII]$. Therefore, although $T_e[OIII]>T(O^{++})$ may still occur at high metallicities, its impact on the total oxygen abundance is expected to be limited, since $O^+$ becomes the dominant oxygen ion in this regime. Our approach is to adopt $T_e[OII] = T_e[SIII]$. This assumption is supported by recent studies reporting tight correlations between $T_e[SIII]$ and $T_e[NII]$, as well as between $T_e[NII]$ and $T_e[OII]$ \citep{Croxall2016,Berg2020,Bresolin2004}. Furthermore, the fact that most of the sulphur is expected to be in the form of $S^{++}$, regardless of the metallicity of the ionised nebula, provides additional support for this approximation \citep[e.g.][]{Garnett1992}.

\begin{figure}[t!]
\centering
\includegraphics[width=0.42\textwidth]{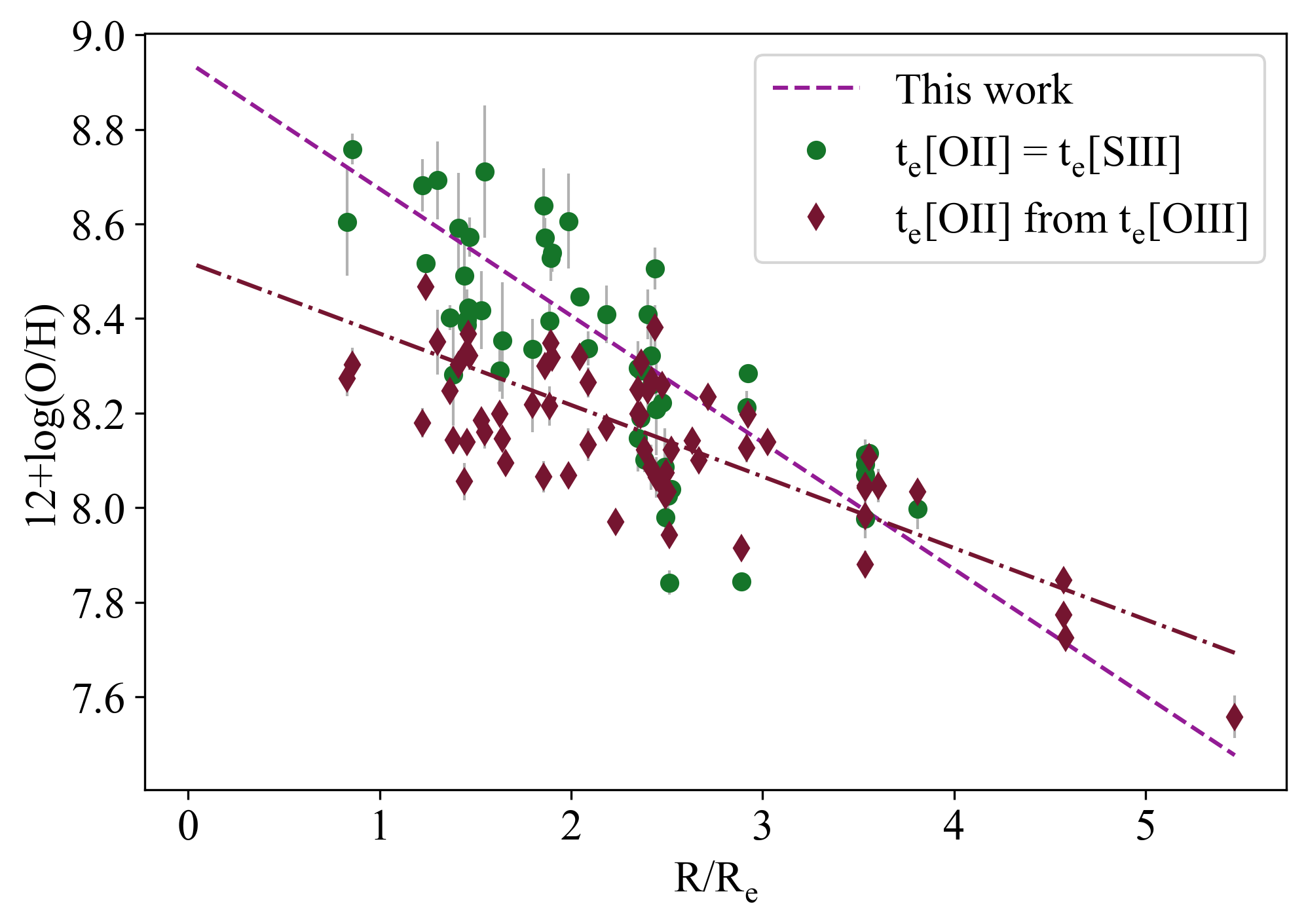} 
\caption{Oxygen abundance gradients derived using the direct $t_e[OIII]$ measurements from this work and adopting two different ionisation structures. The gradients obtained assuming $t_e[OII] = t_e[SIII]$ (this work) and calculating $t_e[OII]$ from $t_e[OIII]$ using the equation provided by \citet{Stasinska1990} are shown as green circles and red diamonds, respectively. The corresponding radial fits to the results calculating $t_e[OII]$ from $t_e[OIII]$ is plotted in the same colour. }
\label{fig_structure_ion}
\end{figure}

Figure \ref{fig_structure_ion} compares the oxygen abundance gradients derived using the temperature structure adopted in this work, $T_e[SIII] = T_e[OII]$, with those obtained using the commonly adopted one, where $T_e[OII]$ is estimated from $T_e[OIII]$. The comparison clearly shows that the commonly adopted temperature structure leads to an overestimation of the temperature in the $O^+$  region, even when $T_e[OIII]$ is directly measured from the auroral lines. This results in an underestimation of the $O^+/H^+$ abundance by up to 0.8 dex, which is the dominant contributor to the total oxygen abundance in the inner regions of the radial gradient. Consequently, adopting this temperature structure results in a shallower oxygen abundance gradient, with a difference of 0.3 dex at the effective radius.

\section{Summary and conclusions} \label{sec:conclusions}\label{conclusions}

Deriving reliable chemical abundances across the radial gradients of spiral galaxies remains challenging because of the wide range of physical conditions within their HII regions. Even direct abundance determinations based on auroral-line electron temperatures are affected by significant biases, particularly in the central, metal-rich regions of galaxies, which are key to constraining abundance gradients. Auroral lines become increasingly faint at low temperatures, hindering their detection, while the exponential dependence of collisionally excited line emissivities on temperature biases the observed emission towards warmer zones of the nebular ionisation structure. This can lead to overestimated electron temperatures and, consequently, underestimated chemical abundances.

In this work, we present an alternative approach to deriving chemical abundances based on the sulphur electron temperature, which we apply to M101 as a test case given its extensive previous study. Our compilation includes regions from \citet{Kennicutt_Garnett1996}, \citet{Kennicutt2003}, \citet{Bresolin2007}, and \citet{Croxall2016}, complemented by 20 SDSS regions from \citet{pilyugin2007}, which we remeasured. The final sample comprises 151 HII regions, 34 within one effective radius, providing extensive coverage of the high-metallicity regime and allowing us to investigate the inner regions of the galaxy in greater detail.

The new proposed methodology relies on deriving $T_e[SIII]$ either from the [SIII] auroral line or, at supersolar metallicities, from the $SO_{23}$ parameter using the calibration provided by \citet{Diaz2007}. Assuming a two-zone ionisation structure, we approximate the low-ionisation zone temperature as $T_e[SIII] \simeq T_e[NII] \simeq T_e[OII]$. The high-ionisation zone temperature, $T_e[OIII]$, is then derived either directly from the oxygen auroral lines or through its empirical relation with $T_e[SIII]$. The main advantages of the new methodology are summarised as follows:

\begin{enumerate}
\item $T_e[SIII]$ can be estimated from strong emission lines even when the auroral line is not detected, reducing the selection effects and associated biases at high metallicities.
\item The use of $T_e[SIII]$ to derive $T_e[OII]$ and $T_e[OIII]$ in high-metallicity regions, where the [OIII] auroral line is often undetectable, mitigates selection effects and the resulting bias towards overestimating electron temperatures and underestimating oxygen abundances.
\item The $SO_{23}$ parameter provides a single-valued relation across the metallicity range, avoiding the uncertainties associated with the branch transition (turnover) that affect other calibrators such as $R_{23}$.
\item Using the calibrator directly to estimate the electron temperature, based on the cooling properties of the nebula, does not require explicit corrections for its ionisation structure or other physical conditions (i.e. $T_{eff}$ or $log(u)$).
\end{enumerate}
Overall, our new method provides a more complete sampling of abundance gradients, particularly in the innermost regions, where metallicities are highest. This not only improves the determination of the gradients themselves but also provides a more robust basis for refining existing emission-line calibrations, such as $R_{23}$ and $O3N2$, over a broader range of metallicities.

\section*{Acknowledgements}
Part of this work has been funded by project Estallidos8 PID2022-136598NB-C33 funded by MCIN/AEI/10.13039/501100011033 and by “ERDF A way of making Europe”.
SZ acknowledges support from the European Union (ERC, WINGS, 101040227). 

\bibliographystyle{mnras}
\bibliography{bibliografia}

\appendix

\section{Additional plots}\label{ap:fig}

\begin{figure}[h!]
\centering
\includegraphics[width=0.42\textwidth]{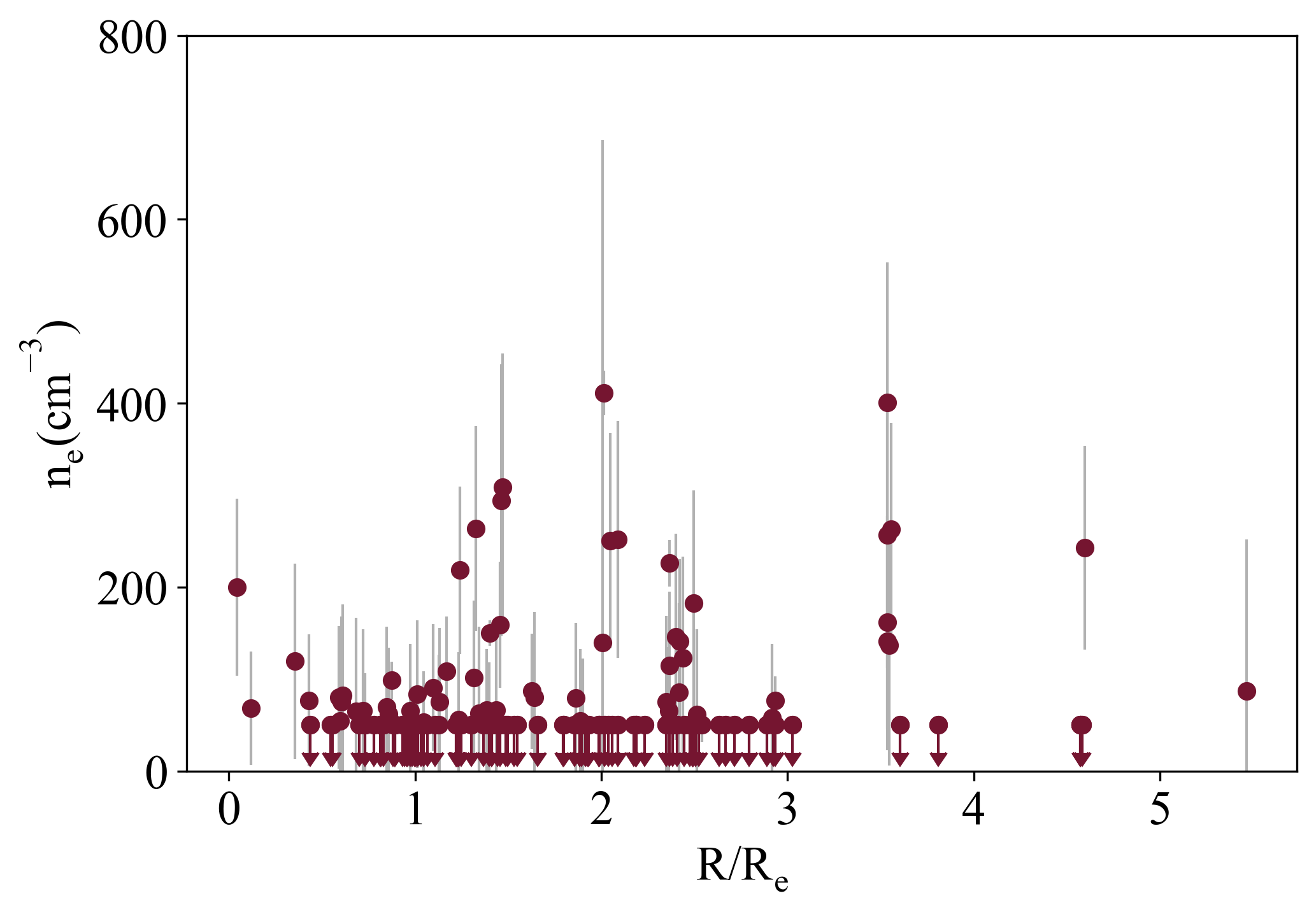}
\caption{Radial gradient of the electron density derived from the $[SII]\lambda6717/\lambda6731$ line ratio. Values at the low-density limit of the diagnostic are treated as upper limits of $n_e=50 cm^{-3}$.}
\label{fig:ne}
\end{figure}

\begin{figure}[h!]
\centering
\includegraphics[width=0.42\textwidth]{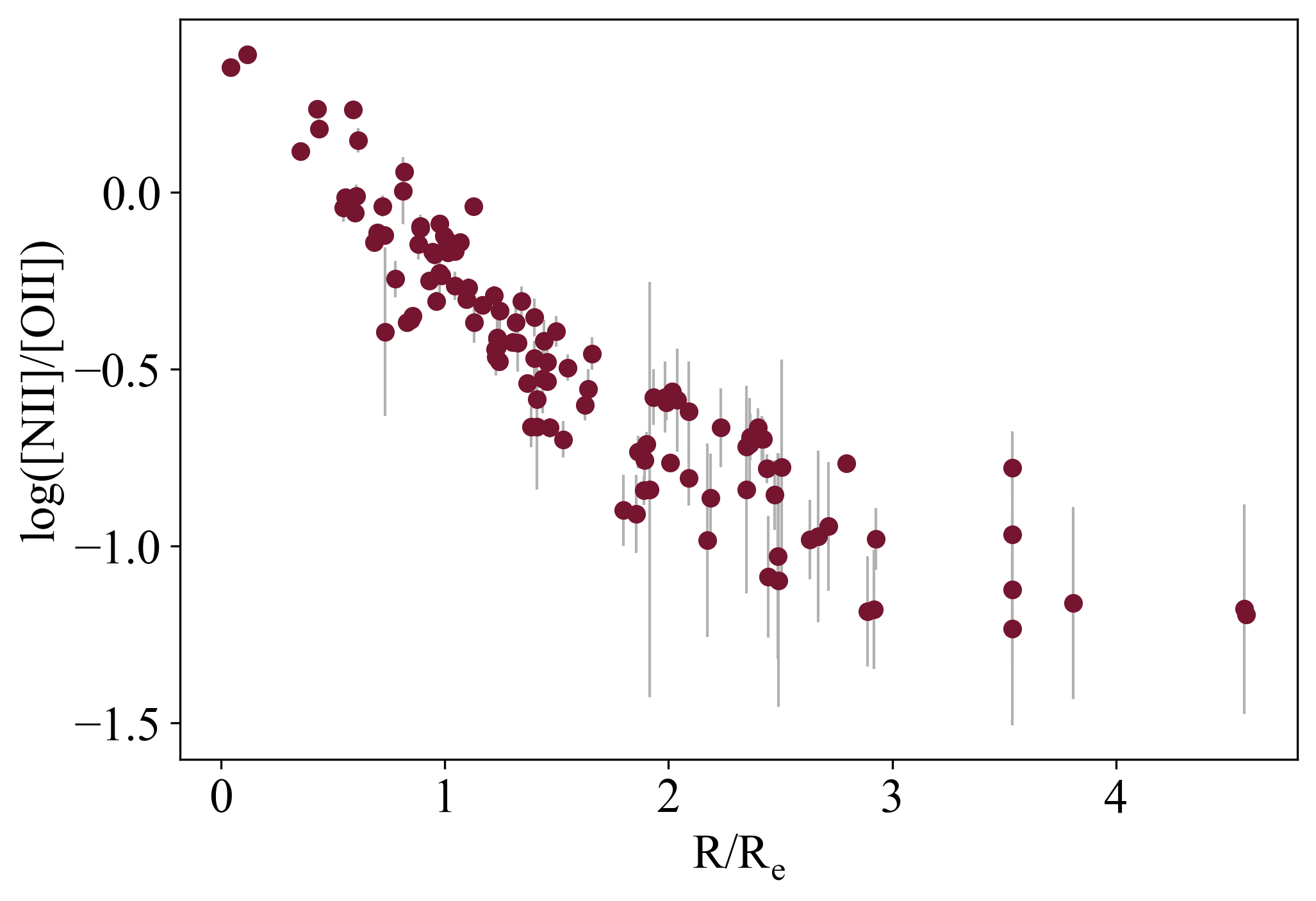}
\caption{Radial gradient of the $[NII]/[OII]$ emission line ratio.}
\label{fig_NIIOII}
\end{figure}

\begin{figure}[h!]
\centering
\includegraphics[width=0.37\textwidth]{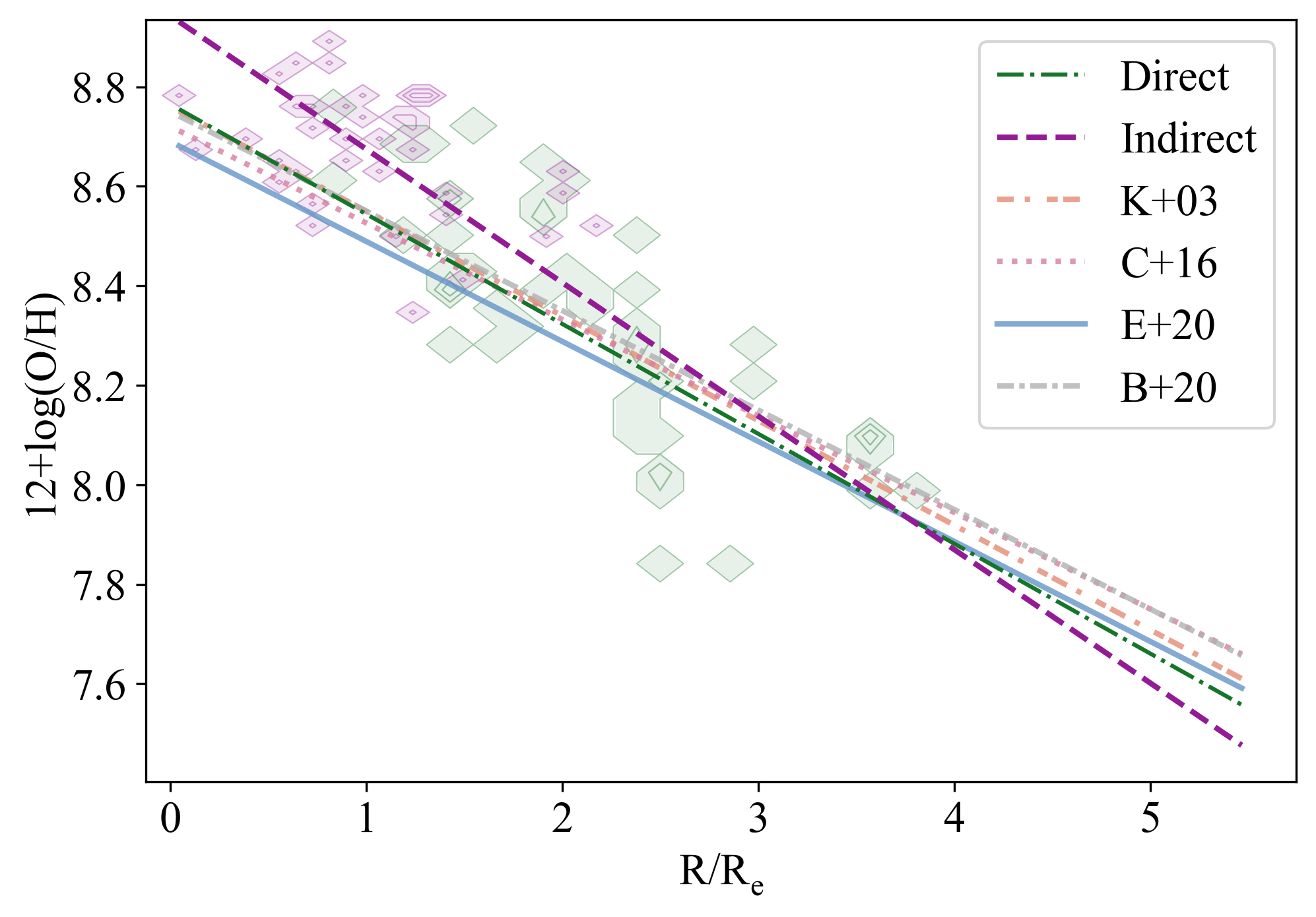}
\includegraphics[width=0.37\textwidth]{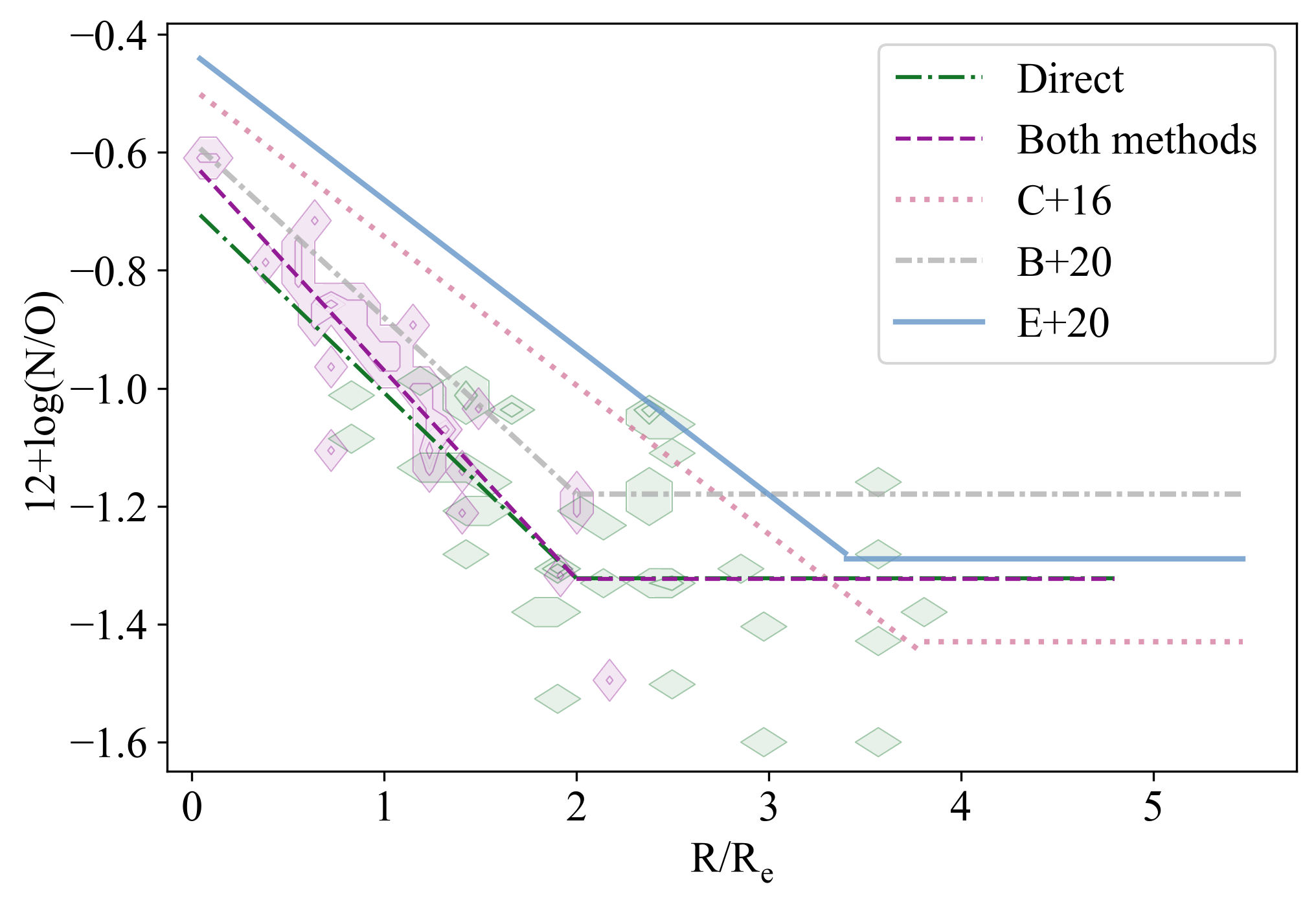}
\caption{Radial abundance gradients of oxygen and nitrogen (top and bottom respectively), derived using the direct and indirect methods. Green and purple contours represent the direct and indirect abundance determinations, respectively, while the corresponding linear fits are shown in the same colours. Previous abundance gradients reported in the literature are also shown \citep{ken03,Croxall2016,Esteban2020,Berg2020}, labelled K+03, C+16, E+20, and B+20, respectively.}
\label{fig_grad_literature}
\end{figure}

\begin{figure}[h!]
\centering
\includegraphics[width=0.37\textwidth]{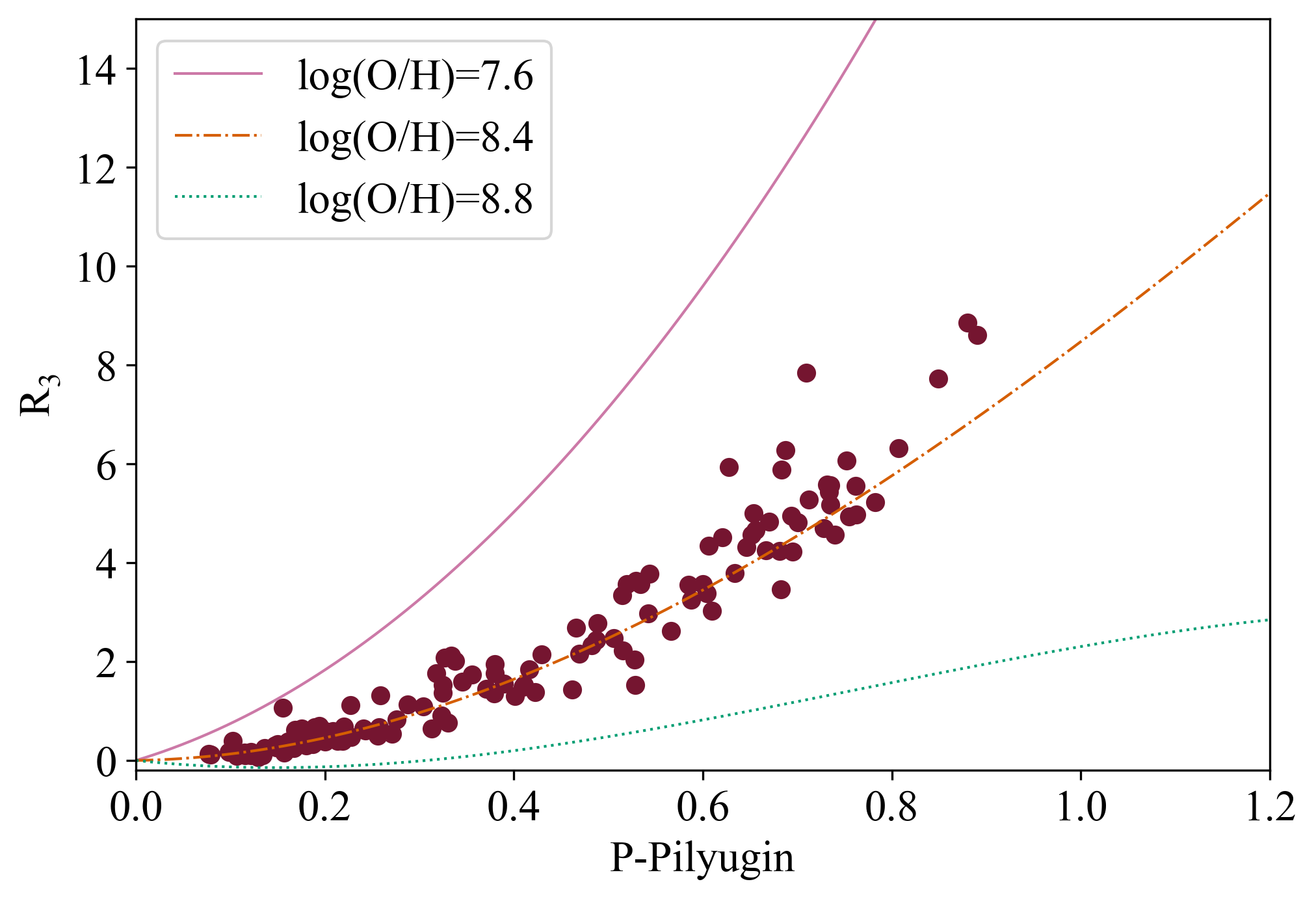}
\includegraphics[width=0.37\textwidth]{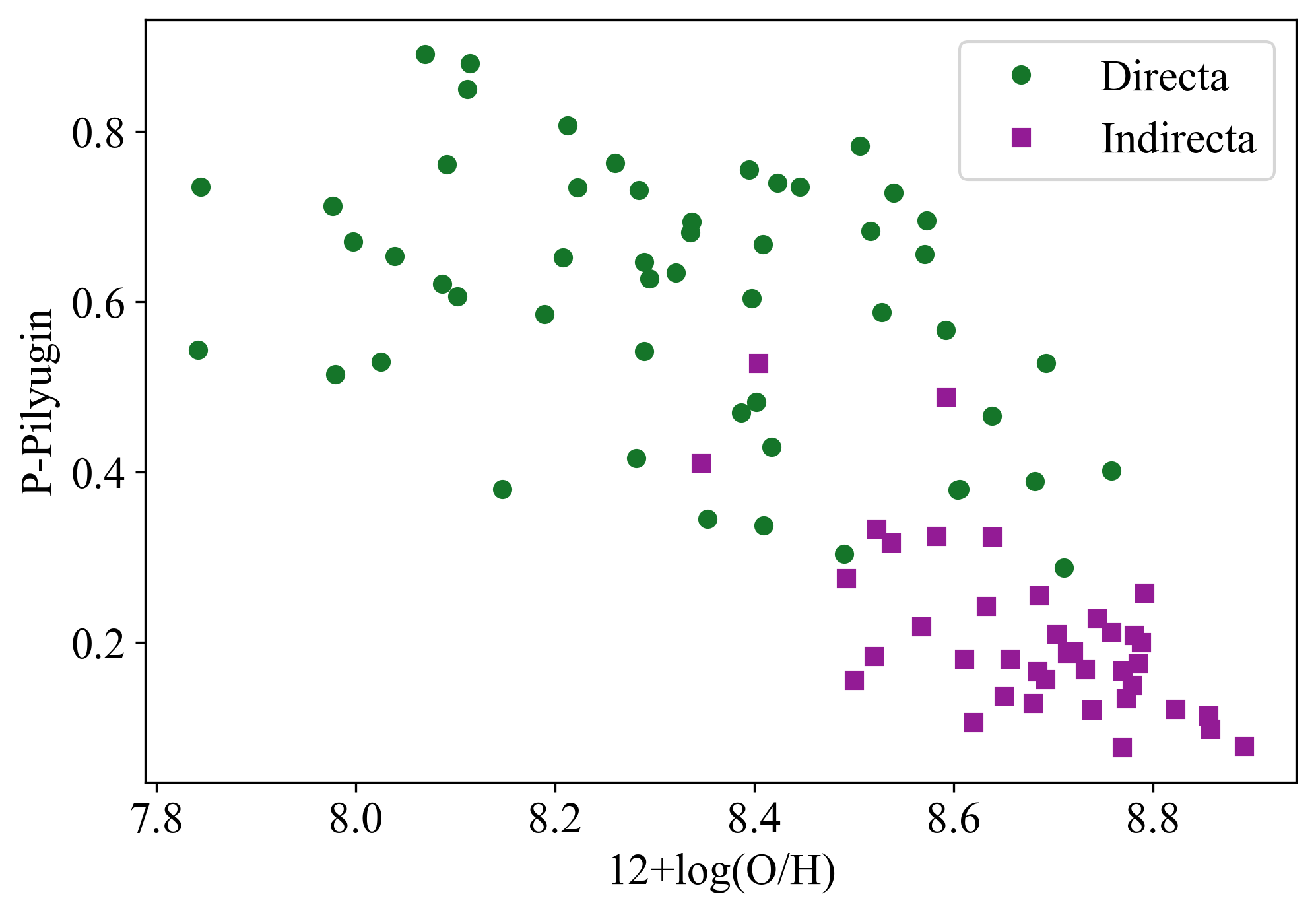}
\caption{Top panel: $R_3$–$P$ diagram with the different parametrisations proposed by \citet{Pilyugin2005} for different abundances. Bottom panel: Excitation parameter $P$ as a function of oxygen abundance. Green circles and purple triangles represent direct and indirect abundance determinations, respectively.}
\label{fig_piliyuigin}
\end{figure}

\onecolumn
\section{Reddening and SDSS data} \label{ap:red}
\input{Tables/SDSS_data}
\input{Tables/SDSS_lines}
\input{Tables/SDSS_reddening}

\end{document}

%% file: Tables/Results_gradients.tex
\begin{table*}[h!]
\centering
\caption{Radial abundance gradients for oxygen, sulfur, and nitrogen, fitted as $12+\log(X/H)=a \cdot R/R_e+b$, and shown in Figure \ref{fig_grad_elementos}.}
    \label{tab:results_gradients}
\begin{tabular}{c|cc|cc|}
\cline{2-5}
        & \multicolumn{2}{c|}{\makebox[5cm][c]{Direct}} & \multicolumn{2}{c|}{\makebox[5cm][c]{Both methods}} \\ \cline{2-5}
        & \multicolumn{1}{c|}{\makebox[2.5cm][c]{a}}  & b & 
        \multicolumn{1}{c|}{\makebox[2.5cm][c]{a}} & b    \\ \hline \hline
\multicolumn{1}{|c|}{Oxygen}  & \multicolumn{1}{c|}{\makebox[2.5cm][c]{-0.221
$\pm$ 0.030}} &  8.765 $\pm$ 0.075 & 

  \multicolumn{1}{c|}{\makebox[2.5cm][c]{-0.268 $\pm$ 0.018}}    &   8.943 $\pm$ 0.031   \\ \hline
\multicolumn{1}{|c|}{Nitrogen} & 
\multicolumn{1}{c|}{\makebox[2.5cm][c]{-0.332 $\pm$ 0.028}}  & 7.889 $\pm$ 0.050
& 
\multicolumn{1}{c|}{\makebox[2.5cm][c]{-0.447 $\pm$ 0.025}}      &   8.181 $\pm$ 0.028   \\ \hline
\end{tabular}
\end{table*}

%% file: Tables/SDSS_data.tex
\begin{table*}[h]
\centering
\small
\begin{tabular}{cccccc}
\hline
ID  & Object & Plate & Mjd   & fiberid & R$_G$ \\ \hline
R1  & \href{http://cas.sdss.org/dr7/en/tools/quicklook/quickobj.asp?plate=1325-52762-345\&mjd=\&fiber=}{SDSS   J140331.32+542114.6}   & 1325  & 52762 & 345     & 0.20  \\
R2  & \href{http://cas.sdss.org/dr7/en/tools/quicklook/quickobj.asp?plate=\%201325\&mjd=52762\&fiber=356}{SDSS   J140339.85+541856.9} & 1325  & 52762 & 356     & 0.32  \\
R3  & \href{http://cas.sdss.org/dr7/en/tools/quicklook/quickobj.asp?plate=1323\&mjd=52797\&fiber=011}{SDSS   J140341.53+541909.3}     & 1323  & 52797 & 11      & 0.33  \\
R4  & \href{http://cas.sdss.org/dr7/en/tools/quicklook/quickobj.asp?plate=\%201325\&mjd=52762\&fiber=348}{SDSS   J140332.34+541719.5} & 1325  & 52762 & 348     & 0.34  \\
R5  & \href{http://cas.sdss.org/dr7/en/tools/quicklook/quickobj.asp?plate=1324\&mjd=53088\&fiber=236}{SDSS   J140301.81+541514.9}     & 1324  & 53088 & 236     & 0.41  \\
R6  & \href{http://cas.sdss.org/dr7/en/tools/quicklook/quickobj.asp?plate=1325\&mjd=52762\&fiber=352}{SDSS   J140353.10+542206.5}     & 1325  & 52762 & 352     & 0.43  \\
R7  & \href{http://cas.sdss.org/dr7/en/tools/quicklook/quickobj.asp?plate=1323\&mjd=52797\&fiber=9}{SDSS   J140253.70+541522.7}       & 1323  & 52797 & 9       & 0.43  \\
R8  & \href{http://cas.sdss.org/dr7/en/tools/quicklook/quickobj.asp?plate=1323\&mjd=52797\&fiber=13}{SDSS   J140354.18+542209.5}      & 1323  & 52797 & 13      & 0.44  \\
R9  & \href{http://cas.sdss.org/dr7/en/tools/quicklook/quickobj.asp?plate=1323\&mjd=52797\&fiber=2}{SDSS   J140301.17+541429.3}       & 1323  & 52797 & 2       & 0.46  \\
R10 & \href{http://cas.sdss.org/dr7/en/tools/quicklook/quickobj.asp?plate=1324\&mjd=53088\&fiber=221}{SDSS   J140357.27+542228.6}     & 1324  & 53088 & 221     & 0.47  \\
R11 & \href{http://cas.sdss.org/dr7/en/tools/quicklook/quickobj.asp?plate=1323\&mjd=52797\&fiber=1}{SDSS   J140246.91+541449.9}                           & 1323  & 52797 & 1       & 0.50  \\
R12 & \href{http://cas.sdss.org/dr7/en/tools/quicklook/quickobj.asp?plate=1324\&mjd=53088\&fiber=223}{SDSS   J140233.07+542530.5}     & 1324  & 53088 & 223     & 0.54  \\
R13 & \href{http://cas.sdss.org/dr7/en/tools/quicklook/quickobj.asp?plate=1323\&mjd=52797\&fiber=8}{SDSS   J113459.47+002509.1}       & 1323  & 52797 & 8       & 0.54  \\
R14 & \href{http://cas.sdss.org/dr7/en/tools/quicklook/quickobj.asp?plate=1324\&mjd=53088\&fiber=271}{SDSS   J140228.20+541633.1}     & 1324  & 53088 & 271     & 0.54  \\
R15 & \href{http://cas.sdss.org/dr7/en/tools/quicklook/quickobj.asp?plate=1324\&mjd=53088\&fiber=279}{SDSS   J140220.20+542002.6}     & 1324  & 53088 & 279     & 0.55  \\
R16 & \href{http://cas.sdss.org/dr7/en/tools/quicklook/quickobj.asp?plate=1324\&mjd=53088\&fiber=263}{SDSS   J140220.61+541746.3}     & 1324  & 53088 & 263     & 0.58  \\
R17 & \href{http://cas.sdss.org/dr7/en/tools/quicklook/quickobj.asp?plate=1325\&mjd=52762\&fiber=350}{SDSS   J140411.23+542518.7}     & 1325  & 52762 & 350     & 0.67  \\
R18 & \href{http://cas.sdss.org/dr7/en/tools/quicklook/quickobj.asp?plate=1323\&mjd=52797\&fiber=16}{SDSS   J140410.98+542521.8}      & 1323  & 52797 & 16      & 0.67  \\
R19 & \href{http://cas.sdss.org/dr7/en/tools/quicklook/quickobj.asp?plate=1324\&mjd=53088\&fiber=234}{SDSS   J140428.62+542352.8}     & 1324  & 53088 & 234     & 0.81  \\
R20 & \href{http://cas.sdss.org/dr7/en/tools/quicklook/quickobj.asp?plate=1323\&mjd=52797\&fiber=66}{SDSS   J140313.72+543543.2}      & 1323  & 52797 & 66      & 1.05 \\
\hline
\end{tabular}
\end{table*}

%% file: Tables/SDSS_lines.tex
\begin{table*}[h]
\centering
\small
\caption{Emission-line fluxes measured normalised by $H\beta$.}
\label{tab:seleccion}
\begin{tabular}{cccccccccc}
\hline
Line & [OIII] & [OIII] &  [OIII] & [SIII] & [NII] & [NII] & [SII] & [SII] & [SIII]\\
$\lambda$ & 4363 & 4959 & 5007  & 6312 & 6548 & 6584 & 6717 & 6731 & 9069 \\ \hline
\multicolumn{1}{c}{Region ID} & \multicolumn{9}{c}{I($\lambda $)$^a$} \\ \hline

R1 & - & 52.3 $\pm$ 0.7 & 147.5 $\pm$ 1.1 & 1.2 $\pm$ 0.0 & 35.6 $\pm$ 0.3 & 84.8 $\pm$ 0.5 & 18.7 $\pm$ 0.2 & 13.8 $\pm$ 0.1 & 35.5 $\pm$ 0.3\\ 
R2 & - & 48.6 $\pm$ 0.6 & 128.9 $\pm$ 1.0 & 0.9 $\pm$ 0.1 & - & - & 23.2 $\pm$ 0.1 & 17.8 $\pm$ 0.1 & 22.1 $\pm$ 0.4\\ 
R3 & - & 77.3 $\pm$ 1.3 & 221.8 $\pm$ 2.7 & 1.3 $\pm$ 0.2 & 29.9 $\pm$ 0.4 & 98.1 $\pm$ 1.0 & - & - & 55.5 $\pm$ 1.2\\ 
R4 & 2.9 $\pm$ 0.7 & 172.1 $\pm$ 4.2 & 474.2 $\pm$ 2.7 & 1.1 $\pm$ 0.1 & 9.0 $\pm$ 0.1 & 28.9 $\pm$ 0.2 & 21.2 $\pm$ 0.2 & 13.3 $\pm$ 0.1 & 15.3 $\pm$ 1.0\\ 
R5 & 3.1 $\pm$ 1.8 & 93.9 $\pm$ 2.6 & 267.0 $\pm$ 5.8 & - & 16.7 $\pm$ 0.6 & 54.5 $\pm$ 1.2 & 42.5 $\pm$ 1.1 & 27.5 $\pm$ 0.8 & 17.3 $\pm$ 2.7\\ 
R6 & 2.6 $\pm$ 0.5 & 144.3 $\pm$ 5.7 & 412.8 $\pm$ 0.8 & 1.6 $\pm$ 0.1 & 8.2 $\pm$ 0.1 & 26.3 $\pm$ 0.2 & 18.1 $\pm$ 0.1 & 12.8 $\pm$ 0.1 & 21.8 $\pm$ 0.3\\ 
R7 & 1.7 $\pm$ 1.0 & 108.1 $\pm$ 2.8 & - & 1.0 $\pm$ 0.2 & 8.7 $\pm$ 0.3 & 25.6 $\pm$ 0.4 & 13.6 $\pm$ 0.3 & 8.6 $\pm$ 0.3 & 19.1 $\pm$ 2.5\\ 
R8 & 5.6 $\pm$ 2.3 & 288.5 $\pm$ 31.0 & - & - & 19.6 $\pm$ 0.3 & 49.4 $\pm$ 0.4 & 51.7 $\pm$ 0.5 & 34.9 $\pm$ 0.4 & 15.4 $\pm$ 1.3\\ 
R9 & 2.5 $\pm$ 0.5 & 164.6 $\pm$ 1.3 & - & 1.6 $\pm$ 0.1 & 10.4 $\pm$ 0.1 & 32.7 $\pm$ 0.2 & 20.1 $\pm$ 0.2 & 17.8 $\pm$ 0.2 & -\\ 
R10 & 1.8 $\pm$ 0.5 & 89.1 $\pm$ 1.1 & 262.0 $\pm$ 1.4 & 1.1 $\pm$ 0.1 & 5.6 $\pm$ 0.1 & 17.8 $\pm$ 0.2 & 14.7 $\pm$ 0.2 & 10.1 $\pm$ 0.2 & 22.2 $\pm$ 1.0\\ 
R11 & 0.8 $\pm$ 0.8 & 81.5 $\pm$ 1.9 & 235.7 $\pm$ 1.7 & 1.0 $\pm$ 0.2 & 11.4 $\pm$ 0.2 & 38.2 $\pm$ 0.4 & 23.2 $\pm$ 0.3 & 15.5 $\pm$ 0.3 & 18.5 $\pm$ 2.1\\ 
R12 & 6.2 $\pm$ 0.7 & 200.1 $\pm$ 2.0 & 612.8 $\pm$ 4.6 & 1.3 $\pm$ 0.2 & 3.0 $\pm$ 0.1 & 9.5 $\pm$ 0.2 & 10.4 $\pm$ 0.2 & 7.0 $\pm$ 0.2 & 22.8 $\pm$ 1.2\\ 
R13 & 1.8 $\pm$ 0.6 & 167.0 $\pm$ 1.9 & 464.3 $\pm$ 3.5 & 2.1 $\pm$ 0.1 & 15.9 $\pm$ 0.2 & 51.0 $\pm$ 0.4 & 30.8 $\pm$ 0.3 & 24.7 $\pm$ 0.3 & 39.7 $\pm$ 0.5\\ 
R15 & 4.0 $\pm$ 0.5 & 144.1 $\pm$ 1.3 & 423.6 $\pm$ 2.3 & 2.3 $\pm$ 0.1 & 4.8 $\pm$ 0.1 & 14.0 $\pm$ 0.2 & 12.7 $\pm$ 0.1 & 8.7 $\pm$ 0.1 & 16.3 $\pm$ 0.6\\ 
R16 & 6.6 $\pm$ 0.6 & 191.4 $\pm$ 1.8 & 534.4 $\pm$ 1.4 & 1.6 $\pm$ 0.1 & 4.1 $\pm$ 0.1 & 11.7 $\pm$ 0.1 & 13.5 $\pm$ 0.1 & 9.7 $\pm$ 0.1 & 10.7 $\pm$ 0.4\\ 
R17 & 4.7 $\pm$ 0.5 & 165.2 $\pm$ 1.6 & - & 1.5 $\pm$ 0.1 & 3.3 $\pm$ 0.1 & 10.4 $\pm$ 0.1 & 11.2 $\pm$ 0.1 & 7.6 $\pm$ 0.1 & 19.0 $\pm$ 0.5\\ 
R18 & 3.1 $\pm$ 0.5 & 131.3 $\pm$ 1.1 & 375.4 $\pm$ 4.2 & 2.0 $\pm$ 0.2 & 5.3 $\pm$ 0.1 & 17.8 $\pm$ 0.2 & 19.3 $\pm$ 0.2 & 14.1 $\pm$ 0.2 & 17.5 $\pm$ 0.7\\ 
R19 & 5.7 $\pm$ 0.4 & 149.6 $\pm$ 2.2 & 441.5 $\pm$ 4.9 & 1.5 $\pm$ 0.1 & 8.3 $\pm$ 0.6 & 18.8 $\pm$ 0.9 & 13.9 $\pm$ 0.8 & 10.5 $\pm$ 0.7 & 14.5 $\pm$ 0.2\\ 
R20 & 18.3 $\pm$ 2.2 & 249.2 $\pm$ 5.7 & 674.9 $\pm$ 13.4 & - & 5.6 $\pm$ 0.5 & 9.7 $\pm$ 0.5 & 11.5 $\pm$ 0.5 & 9.3 $\pm$ 0.5 & 5.9 $\pm$ 0.5\\ 

\hline                                
\end{tabular}
\begin{tablenotes}
\centering
\item $^a$ Values normalised to I(H$\beta$) 10$^{-2}$. 
\end{tablenotes}
\end{table*}

%% file: Tables/SDSS_reddening.tex
\begin{table*}[h!]
\centering
\caption{Emission-line fluxes measured in units of $10^{-17}$ 10$^{-17}$ erg/s/cm$^2$/\AA, and $c(H\beta)$ derived from the fits.}
\label{tab:seleccion}
\begin{tabular}{cccccc}
\hline \hline
Region ID & c(H$\beta$) & H9 & H$\delta$ & H$\gamma$ & H$\beta$  \\ \hline
R1 & 0.5103 $\pm$ 0.0011 & 409.1 $\pm$ 16.7 & 1717.0 $\pm$ 27.6 & 3247.3 $\pm$ 15.7 & 7911.4 $\pm$ 34.8\\ 
R2 & 0.8314 $\pm$ 0.0009 & 149.2 $\pm$ 7.3 & 609.6 $\pm$ 11.7 & 1265.2 $\pm$ 10.6 & 3215.3 $\pm$ 15.5\\ 
R3 & 0.3096 $\pm$ 0.0004 & 153.5 $\pm$ 12.5 & 565.2 $\pm$ 22.3 & 1070.3 $\pm$ 19.0 & 2512.5 $\pm$ 23.4\\ 
R4 & 0.7342 $\pm$ 0.0297 & 125.4 $\pm$ 7.5 & 527.3 $\pm$ 11.2 & 1044.1 $\pm$ 8.0 & 2429.0 $\pm$ 10.9\\ 
R5 & 0.2590 $\pm$ 0.0029 & - & 65.4 $\pm$ 5.8 & 124.8 $\pm$ 5.8 & 283.4 $\pm$ 4.9\\ 
R6 & 0.4591 $\pm$ 0.0001 & 231.7 $\pm$ 12.9 & 892.5 $\pm$ 25.1 & 1735.5 $\pm$ 14.5 & 4221.9 $\pm$ 32.4\\ 
R7 & 0.7059 $\pm$ 0.0155 & 53.1 $\pm$ 8.7 & 250.7 $\pm$ 18.6 & 406.6 $\pm$ 18.0 & 947.5 $\pm$ 19.5\\ 
R8 & 1.2129 $\pm$ 0.1377 & 228.8 $\pm$ 14.6 & 827.6 $\pm$ 25.3 & 1493.4 $\pm$ 18.2 & -\\ 
R9 & 0.3060 $\pm$ 0.0029 & - & 3841.1 $\pm$ 97.6 & 6696.6 $\pm$ 36.7 & 15844.0 $\pm$ 81.9\\ 
R10 & 0.3297 $\pm$ 0.0009 & - & 287.7 $\pm$ 10.4 & 551.4 $\pm$ 9.1 & 1257.1 $\pm$ 10.0\\ 
R11 & 0.5732 $\pm$ 0.0230 & 104.9 $\pm$ 11.3 & 388.8 $\pm$ 18.1 & 722.8 $\pm$ 20.6 & 1486.7 $\pm$ 20.1\\ 
R12 & 0.1111 $\pm$ 0.0001 & - & 255.0 $\pm$ 10.6 & 472.1 $\pm$ 13.2 & 1054.9 $\pm$ 11.0\\ 
R13 & 0.4572 $\pm$ 0.0038 & 397.4 $\pm$ 23.2 & 1492.4 $\pm$ 45.4 & 2717.0 $\pm$ 25.7 & 5970.2 $\pm$ 40.5\\ 
R14 & - & 194.3 $\pm$ 10.3 & 666.7 $\pm$ 15.7 & 1265.1 $\pm$ 12.6 & 2778.1 $\pm$ 14.7\\ 
R15 & 0.2859 $\pm$ 0.0049 & 120.2 $\pm$ 7.6 & 504.0 $\pm$ 13.1 & 939.6 $\pm$ 10.9 & 2014.3 $\pm$ 12.9\\ 
R16 & 0.3628 $\pm$ 0.0015 & 148.0 $\pm$ 9.3 & 600.2 $\pm$ 18.4 & 1145.6 $\pm$ 16.9 & 2557.3 $\pm$ 18.9\\ 
R17 & 0.2671 $\pm$ 0.0036 & 376.9 $\pm$ 15.6 & 1345.7 $\pm$ 27.0 & 2457.3 $\pm$ 14.1 & 5209.0 $\pm$ 27.5\\ 
R18 & 0.1452 $\pm$ 0.0002 & 130.4 $\pm$ 10.0 & - & - & 1975.0 $\pm$ 14.7\\ 
R19 & 0.1164 $\pm$ 0.0008 & 528.0 $\pm$ 24.6 & 1769.6 $\pm$ 39.3 & 3243.9 $\pm$ 19.4 & 7122.1 $\pm$ 44.2\\ 
R20 & 0.8561 $\pm$ 0.0092 & 27.6 $\pm$ 6.5 & - & - & 605.5 $\pm$ 9.4\\ 

\end{tabular}

\begin{tabular}{ccccccc}
\hline \hline
Region ID & HeI$\lambda$4922 & HeI$\lambda$5876 & H$\alpha$ & HeI$\lambda$6678& HeI$\lambda$7281 & Pa11\\ \hline
R1 & 88.9 $\pm$ 16.1 & 1061.9 $\pm$ 6.6 & - & 379.2 $\pm$ 7.6 & 83.7 $\pm$ 13.0 & 1061.9 $\pm$ 6.6\\ 
R2 & 31.4 $\pm$ 7.7 & 442.6 $\pm$ 4.5 & 16099.4 $\pm$ 27.7 & 169.9 $\pm$ 3.7 & 34.5 $\pm$ 10.8 & 442.6 $\pm$ 4.5\\
R3 & - & 429.3 $\pm$ 8.9 & - & 136.8 $\pm$ 4.4 & 26.6 $\pm$ 9.9 & 429.3 $\pm$ 8.9\\ 
R4 & 15.7 $\pm$ 5.7 & 298.6 $\pm$ 4.2 & - & 80.8 $\pm$ 2.6 & - & 298.6 $\pm$ 4.2\\ 
R5 & - & 40.3 $\pm$ 2.7 & - & 12.7 $\pm$ 1.5 & - & 40.3 $\pm$ 2.7\\ 
R6 & 45.9 $\pm$ 12.7 & 569.9 $\pm$ 7.6 & 16480.2 $\pm$ 29.0 & 183.2 $\pm$ 3.5 & - & 569.9 $\pm$ 7.6\\ 
R7 & - & 128.4 $\pm$ 6.6 & 3688.4 $\pm$ 17.4 & 50.0 $\pm$ 3.1 & - & 128.4 $\pm$ 6.6\\ 
R8 & 13.5 $\pm$ 7.2 & 293.5 $\pm$ 7.3 & 4839.9 $\pm$ 14.3 & 90.7 $\pm$ 3.0 & - & 293.5 $\pm$ 7.3\\ 
R9 & - & 1914.9 $\pm$ 14.4 & - & 623.4 $\pm$ 11.1 & 144.1 $\pm$ 50.5 & 1914.9 $\pm$ 14.4\\ 
R10 & - & 174.1 $\pm$ 4.2 & 4630.9 $\pm$ 13.3 & 56.1 $\pm$ 2.0 & - & 174.1 $\pm$ 4.2\\ 
R11 & - & 152.5 $\pm$ 6.2 & 4966.7 $\pm$ 17.7 & 57.3 $\pm$ 3.1 & - & 152.5 $\pm$ 6.2\\ 
R12 & - & - & 3220.0 $\pm$ 10.8 & 40.2 $\pm$ 2.0 & - & -\\ 
R13 & 71.6 $\pm$ 15.0 & 835.0 $\pm$ 10.2 & - & 294.0 $\pm$ 6.7 & - & 835.0 $\pm$ 10.2\\ 
R14 & - & 298.3 $\pm$ 4.0 & - & 105.7 $\pm$ 2.6 & 16.9 $\pm$ 6.7 & 298.3 $\pm$ 4.0\\ 
R15 & 20.4 $\pm$ 5.4 & 222.8 $\pm$ 3.9 & 6276.2 $\pm$ 13.2 & 72.2 $\pm$ 2.1 & 14.5 $\pm$ 5.0 & 222.8 $\pm$ 3.9\\ 
R16 & - & 282.5 $\pm$ 4.8 & 8699.9 $\pm$ 17.0 & 94.5 $\pm$ 2.4 & 17.7 $\pm$ 6.6 & 282.5 $\pm$ 4.8\\ 
R17 & 51.5 $\pm$ 13.8 & 552.0 $\pm$ 5.0 & 15830.6 $\pm$ 30.2 & 180.4 $\pm$ 4.2 & - & 552.0 $\pm$ 5.0\\ 
R18 & - & 244.6 $\pm$ 6.5 & 6153.2 $\pm$ 15.9 & 73.7 $\pm$ 2.8 & - & 244.6 $\pm$ 6.5\\ 
R19 & 68.5 $\pm$ 20.4 & 718.2 $\pm$ 5.8 & 22797.0 $\pm$ 253.2 & 218.9 $\pm$ 38.1 & 41.8 $\pm$ 19.2 & 718.2 $\pm$ 5.8\\ 
R20 & - & 59.2 $\pm$ 4.5 & - & 22.8 $\pm$ 2.4 & - & 59.2 $\pm$ 4.5\\ \hline
\end{tabular}
\end{table*}